# Inflation and the Higgs Scalar

DANIEL GREEN
*Fermilab, P.O. Box 500*
*Batavia, IL 60510, USA*
dgreen@fnal.gov

## Abstract

This note makes a self-contained exposition of the basic facts of big bang cosmology as they relate to inflation. The fundamental problems with that model are then explored. A simple scalar model of inflation is evaluated which provides the solution of those problems and makes predictions which will soon be definitively tested. The possibility that the recently discovered fundamental Higgs scalar field drives inflation is explored.

## 1. Introduction

The last few years have yielded remarkable discoveries in physics. In particle physics it appears that a fundamental scalar field exists [1], [2]. The Higgs boson is the excitation of that field and is measured to have a mass of about 126 GeV and to have spin zero and positive parity. The Higgs field is the first fundamental scalar to be discovered in physics.

The Cosmic Microwave Background, (CMB) is the cooled fireball of the Big Bang. It is known to have a uniform temperature to parts per $10^5$ but has [3], [4] well measured temperature perturbations and the associated baryon perturbations are thought to evolve gravitationally and thus to provide the seeds of the current structure of the Universe.

In addition, the Universe appears to contain, at present an unknown "dark energy" [5] which is currently the majority energy density of the Universe, larger than either matter or radiation. This may, indeed, be a fundamental scalar field like the Higgs with a vacuum expectation value which defines a cosmological constant or it may evolve in time as a dynamic field.

"Big Bang" (BB) cosmology is a very successful "standard model" in cosmology [6]. The CMB is the cooled remnant of the BB. The abundances of the light nuclei from an early hot phase are accurately predicted by BB cosmology. However, the model cannot explain the uniformity of the CMB because the CMB consists of many regions not causally connected in the context of the BB model. In addition, the Universe appears to be spatially flat [7]. However in BB cosmology the spatial curvature is not stable so that the initial conditions for BB cosmology would need to be fantastically fine-tuned in order to successfully predict the presently observed small value of the curvature of space-time.

These basic issues for BB cosmology have led to the hypothesis of "inflation" which postulates an unknown scalar field which causes an exponential expansion of the Universe at very early times [8], [9]. This attractive hypothesis can solve the problems of flatness and causal CMB connectivity which are inherent in BB cosmology. In addition, the zero point quantum fluctuations of this postulated field provide a natural explanation of the CMB temperature perturbations and the associated large-scale structure of the Universe.

Researchers are now searching for gravitational waves imprinted on the CMB [10], [11]. These would be strong evidence for inflation since metrical fluctuations are produced in inflationary models. This note is a very basic exposition of the BB cosmology and an inflationary model. The simplest scalar model will be explored in detail first because it is easy to understand, contains all the basic elements of the inflationary model [12], and is presently consistent with the data. Another model is covered briefly for comparison. Finally, the Higgs field can be the inflationary field, although a non-minimal coupling of the Higgs to gravity is needed.

## 2. Units and Constants

The units used in this note are GeV, m, and sec. There is only one coupling constant, the gravitational constant of Newton, $G_N$, which has dimensions in contrast with the dimensionless couplings in particle physics. That constant is subsumed by using the Planck mass to set mass scales. Since inflation deals with very high energies, the Planck mass, $M_p$, is a natural scale. Related scales are the Planck length $L_{pl}$ and the Planck time $t_{pl}$. Natural units are used, $\hbar = c = 1, G_N = 1/M_p^2$, so that, in principle, all quantities could be expressed in energy units.

$$\begin{aligned} M_p &= \sqrt{\hbar c / G_N} \sim 1.2 x 10^{19} GeV \\ L_{pl} &= \sqrt{\hbar G_N / c^3} \sim 1.6 x 10^{-35} m \\ t_{pl} &= L_p / c \sim 5 x 10^{-44} \sec \end{aligned} \quad (1)$$

Several numerical quantities used in this note are quoted from the analyses of the Particle Data Group [14] and appear in Table1. Present quantities are indicated by an o subscript. A basic fact of cosmology is that the Universe is expanding. This is known because the line spectra of distant stars are Doppler shifted into the red indicating an effective recession velocity. The Hubble parameter, H, relates the velocity of recession and the distance of an observed object. It is assumed that matter is essentially at rest in the expanding space. The Hubble distance $D_H$, defines the distance where the recession velocity due to expansion is c and therefore is approximately the observable size of the Universe. The present CMB has cooled off during the expansion and now has a temperature of 2.726 degrees Kelvin and a number density of about 4.11 x $10^8$ photons/m$^3$.

Table 1: Present Values of Selected Cosmological Quantities [14]

| Quantity | Present Value |
|---|---|
| To – CMB Temperature ($^o$K) | 2.726 |
| $t_H$ = 1/ Ho = Hubble Time (sec) | 4.77 x $10^{17}$ = 14.53 Gyr |
| c/Ho = $D_H$ = Hubble Distance (m) | 1.37 x $10^{26}$ |
| $\rho_c$ – critical density (GeV/m$^3$) | 5.6 |
| Ωm – matter fraction (dark matter dominant) | 0.315 |
| Ωr – radiation fraction | 4.6 x $10^{-5}$ |
| Ωv – dark energy or vacuum fraction | 0.685 |

In Table1, c/H$_o$ is the present Hubble distance, $D_H$. The Hubble time is defined to be $t_H$ = 1/H$_o$, or 14.53 billion years. If the Universe were always metrically expanding by a power law, $t^n$, then the Hubble distance would be ct/n and the Hubble time equal to t/n as will be derived later. Detailed analysis of the entire evolution of the Universe yields a value of the present time since the BB of $t_o$ = 13.81 +-0.05 billion years, or approximately the Hubble time [14]. That fact, H$_o$t$_o$ = 0.950, is somewhat coincidental.

The critical density is defined to be the density such that the spatial curvature is zero or "flat". The present value is about 5.6 GeV/m$^3$. The present Universal energy density is about 68% dark energy, 31% matter, both dark and baryonic, and there is a small fraction of the total energy density due to radiation. The total critical density is $\rho_c \sim 10^{-46} GeV^4 = (3.2 x 10^{-12} GeV)^4$. The dark energy (DE) density alone is about $(2.3 \text{ meV})^4$. The vacuum energy density could have either sign. However, the accelerated expansion seen today using supernovas as "standard candles" means that the vacuum energy is repulsive. The scaled density of baryons is 5%, while that of the dark matter (DM) is 26.5%. For the baryons the density is about 1/3 of a proton per cubic meter.

The present Universe is "flat" in that the sum of the ratios of all energy densities to the critical density, $\Omega = \rho/\rho_c$, is 1.001 +- 0.003 [14]. Flatness will be assumed in what follows except when exploring why that is a problem for BB cosmology. Neutrinos are ignored in Table1 because their masses are not known except that they are small. For very light masses they would contribute to the "radiation" , or relativistic particle, energy density. At later times they would cool, become non-relativistic and contribute to the matter energy density. The present limit on the scaled density for neutrinos arises from limits due to the effect of neutrinos on galactic structure formation [14] and is $\Omega_v < 0.0055$.

## 3. Big Bang Cosmology

Assume that the Universe is homogeneous and isotropic [6] which is as simple an assumption as is possible. The space-time metric is then of the Robertson-Walker form. Assume that the Universe is spatially flat from the beginning, in agreement with observation so that the curvature parameter in the metric, k is zero. The curvature will be discussed later. Spatial coordinates such as r are commoving with the metric and dimensionless, participating in the expansion of the Universe and residing at fixed locations. The physical distance scale is set by the single parameter a(t) which modifies the spatial part of the interval of special relativity and tracks the evolution of the Universe. The parameter a(t) completely defines the flat geometry of the Universe.

A static and flat Universe would have a(t) of one. The metrical interval with co-moving spatial coordinates $r, \theta, \phi$ is;

$$ds^2 = (cdt)^2 - a^2(t)(dr^2 + r^2 d\Omega^2) \tag{2}$$

The dynamics of the scale factor a(t) are set by the energy content of the Universe as derived from the Einstein field equations. The energy content defines the metric which is a key element of General Relativity (GR);

$$H^2 = (\dot{a}/a)^2 = (8\pi/3M_p^2)\rho \tag{3}$$

H(t) is the dynamic Hubble parameter and the dot over a(t) denotes a time derivative with respect to t. The density ρ refers to all energy densities; matter, radiation and dark energy or vacuum energy. The Hubble parameter is the fundamental scale for the time evolution of the Universe.

The energy density has a time dependence determined by the equation of state obeyed by the particular type of energy. The basic relationship has to do with the change of energy within a physical volume V which is proportional to a comoving volume times a(t)$^3$. That change, d(ρa(t)$^3$) is due to the pressure, p, doing work, -pdV. Differentiating the energy conservation equation, d(ρa(t)$^3$) + pdV = 0, with V ~ a(t)$^3$, dρ+3(da(t)/a(t))(ρ+p) = 0. The pressure and density act in concert in GR which is not very intuitive. The pressure can do no work since the Universe has no edge. However, the pressure contains energy since it has dimensions of an energy density. The time dependence of the energy density in an expanding space is therefore;

$$\dot{\rho} + 3H(\rho + p) = 0 \tag{4}$$

The two terms define the behavior of the uniform fluid which contains the energy in a dynamic Universe. The H term provides the "friction", while the density term tracks the reduction in density due to the volume increase during expansion and the pressure term tracks the reduction in pressure energy during expansion. For matter domination of the energy density, p is assumed to be zero because the matter is assumed to be non-relativistic, while for radiation domination the classical photon equation of state, p = ρ/3, is assumed.

In the case of temperature, or energy, the scaling as the inverse of a(t) occurs because the fluid cools upon expansion, T ~ 1/a(t). Similarly, the physical wavelength scales as a(t) since waves are stretched or red shifted by the expansion.

The equation for da/dt, Eq.3, can be solved easily in the case where matter or radiation dominates. For matter domination, the density scales with inverse volume or as the inverse cube of the scale a(t), while for radiation it scales as temperature to the fourth power (Stefan-Boltzmann law) or as the inverse fourth power of a(t).

Solving Eq.3 for a(t) in the case of matter domination, the scale a(t) goes as the 2/3 power of t while the energy density goes as the inverse square of t. In the case of radiation dominance a(t) goes as the square root of t while the density again goes as the inverse square of the time t. Therefore in both cases the constancy of $\rho t^2$ holds. Extrapolating to the past, at reduced scale a(t), the radiation is expected to dominate at early times and the temperature will rise. In addition, there is an implied singularity at t of zero, the "Big Bang" giving the Universe a finite age.

The two equations, Eq.3 and Eq.4, can be combined by first differentiating the Hubble expression and then substituting for the time derivative of ρ. The result is the acceleration equation.

$$\ddot{a}/a = -4\pi/(3M_p^2)(\rho + 3p) \tag{5}$$

Clearly, both for matter and for radiation domination, the acceleration is negative; an expanding Universe dominated by the energy density of matter or radiation will decelerate. The acceleration in a matter dominated phase goes as $-1/a(t)^2$ while in a radiation dominated phase it goes as $-1/a(t)^3$. In either case the deceleration is large when a(t) is small and then slows as the scale factor grows.

In the case of dark energy or "vacuum energy" the energy density, $\rho_v$, of this "cosmological term" is constant with respect to a(t) since the density is proportional to the space-time metric itself and tracks the changes in scale. Specifically the pressure is negative and equal in magnitude to the density so that the time dependence of the density is zero, Eq.4. Therefore, H is a constant, Eq.3, and the scale factor grows exponentially in time. The acceleration in this case is positive.

$$da/a = dt\sqrt{8\pi\rho_v/3M_p^2} = Hdt$$
$$\ln(a) = Ht, a \sim e^{Ht}$$
$$\ddot{a} = H^2 a \tag{6}$$

Photons travel on null geodesics, $ds^2 = 0$ in the Robertson-Walker metric as is the case in Special Relativity (SR). In this case, $cdt = a(t)dr$ for a radial photon path and a photon emitted at time $t = 0$ and absorbed at a time t has a "conformal time" $\tau$ which is defined here to be dimensionless.

$$\tau = c\int_0^t dt/a(t) = \int_0^r dr = r \tag{7}$$

This definition of conformal time accounts for the expansion of the Universe during the travel time of the photon and restores the 'light cones' familiar from the Minkowski flat space of SR with the limitation that $\tau > 0$. Light starting from coordinate distance $r = 0$ at $\tau = 0$ arrives at $r = \tau$. Any more distant point is outside the light cone and is not yet visible assuming that the Universe has a beginning at time zero. The conformal clock time slows with respect to the coordinate clock time as the Universe expands.

$$ds^2 = 0 = a^2(t)[d^2\tau - dr^2] \tag{8}$$

The expected power law behaviors of the three types of energy density are easily derivable from the definitions and are shown in Table2 for energy density, scale factor, conformal time, and Hubble parameter. The Universe evolves under a mix of energy densities.

Table 2: Time Dependence of Selected Cosmological Quantities.

|  | $\rho(a)$ | $a(t)$ | $\tau$ | H |
|---|---|---|---|---|
| Matter | $1/a^3$ | $t^{2/3}$ | $t^{1/3}$ | $(2/3)/t$ |
| Radiation | $1/a^4$ | $t^{1/2}$ | $t^{1/2}$ | $(1/2)/t$ |
| Vacuum Energy | constant | $e^{Ht}$ | $c(e^{Ht}-1)/aH$ | constant |

## 4. Horizons

The Hubble law for a dynamical Universe follows from the metric, Eq.2. The recession velocity of galaxies at rest in the space is proportional to physical distance. The relative physical velocity of two points in a spatially flat space with separation R is HR since $R = ar$, $dR/dt = v = rda/dt = HR$. If that velocity is constant then $R = 0$ when $t_H = 1/H_o$. The present particle horizon is the physical Hubble distance $D_H = c/H_o$ which corresponds to that physical distance R where the velocity is c. The comoving Hubble distance is $d_H = c/[H_o a_o]$. In this note the scale is normalized such that $a_o = ct_H \sim ct_o$ which sets $d_H = 1$. Upper case symbols denote physical quantities, while lower case ones denote comoving quantities.

The Hubble distance expands faster than the galaxies since $1/H \sim t$ is greater than $a(t) \sim t^n$ for n less than 1 (Table2) so that more of the Universe is included inside the horizon and is therefore visible as time goes on. Eventually the entire Universe becomes visible. The Hubble distance expands faster than the Universe, overtaking the receding galaxies. In power law situations the Hubble distance has an expansion velocity $cd(1/H)/dt = v_H = c/n$, greater than c. This behavior is true for a decelerating Universe composed of matter and radiation. This is not a physical velocity, violating special relativity, but the velocity of expansion of the metric itself.

The maximum comoving distance, dr, that light can go in time dt is just $d\tau$ (Eq.7). This is the comoving particle horizon while the physical horizon is $a(t)d\tau$. The horizon occurs starting at $\tau = 0$, at proper distance r is $\tau_o$. The relationships of conformal and coordinate time for the three types of energy density are shown in Table2. The comoving particle horizon is $\tau(t)$, assuming $t_i = 0$ for initial time. The horizon at present assuming a single power law behavior is approximately;

$$\tau_o \sim [ct_o / a(t_o)(1-n)] \tag{9}$$

The horizon with $n = 2/3$ for matter dominated evolution results in an estimate for $\tau_o$ of 3. The comoving horizon is $\tau(t_o) - \tau(o) \sim 3.0$ in BB cosmology because no new physics intervenes at small times. The physical horizon distance is $D_{hor} \sim ct/(1-n)$. The Hubble "constant" is $H = n/t$, so that the physical horizon distance is $D_{hor} \sim c/H(t)[n/(1-n)]$ which is twice the Hubble distance in a matter dominated epoch. The physical horizon distance in BB cosmology is then $\sim 3.0(c/H_o) \sim 4.1 \times 10^{26}$ m. The most red-shifted galaxy visible today is at a distance of about $1.92(c/H_o)$ or about 60% of the current horizon distance.

Some comments on the conformal time definition are needed. The definition of $\tau$ depends on the limits of the integral. For radiation or matter domination both the definite and indefinite integrals are the same for limits on t of zero and t leading to Eq.9 and defining $t = 0$ to be the Big Bang. For vacuum energy, the indefinite and definite conformal times differ. Choosing the origin of time to be the end of inflation yields a negative conformal time. Choosing the time origin to be the start of inflation, the conformal time is $\tau = c(e^{Ht} - 1)/[Ha(t)]$ as seen in Table2. This choice for conformal time is made in this note.

The more general question of the communication by light emission and reception is instructive. The coordinate distance between emission and reception of light at $t_e$ and $t_o$ respectively is $\tau(t_o) - \tau(t_e)$. The physical distance at reception today is $D_o = a_o[\tau(t_o) - \tau(t_e)]$ and similarly for the distance at emission. For small differences in the times of emission and reception the distance is about $c(t_o - t_e)$ because the effect of expansion is small. Making a Taylor expansion, the distance $D_o$ is approximately, $D_o \sim c(t_o - t_e) + cH_o(t_o - t_e)^2/2$, where the quadratic deviations due to expansion during the transit time of the light appear.

The full dependence of the physical distance on emission time is shown in Fig.1 where the results of numerically integrating Eq.3 and Eq. 7 with values for the energy densities taken from Table1 were used to find the conformal time as a function of emission time. The distance at reception is larger than $ct_o$ and the distance at emission is smaller. The present horizon at $t_e = 0$

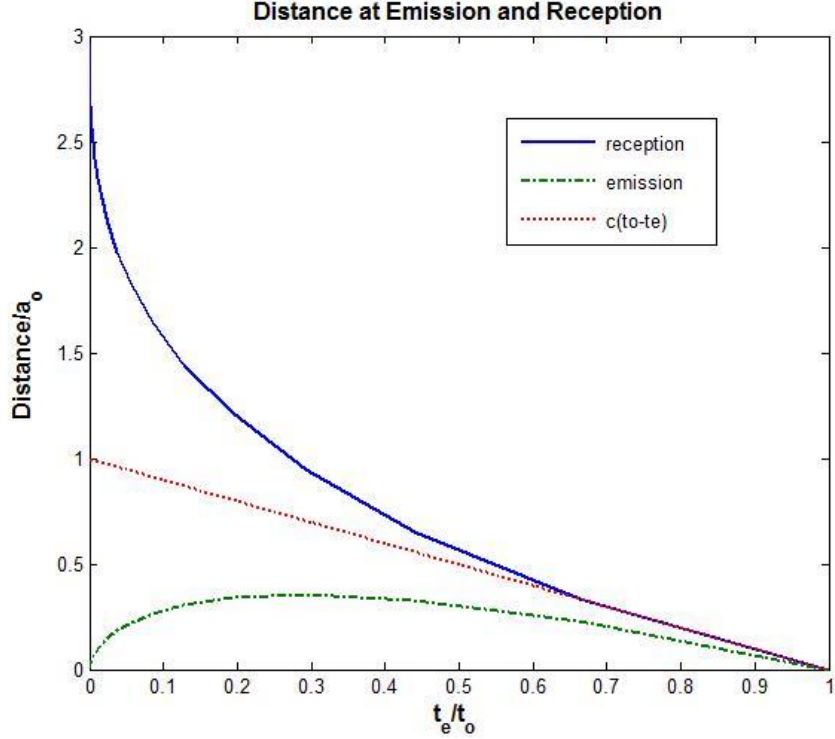

Figure 1: The physical distance at reception and emission, normalized to $a_o$, for a reception time $t_o$, at present, as a function of the scaled emission time $t_e/t_o$.

has a present distance about 3 times the Hubble distance scaled to $a_o$, .

In the case of vacuum energy, space-time expands rapidly which stretches the causal regions to sizes much larger than the Hubble distance which is constant by assumption. Since a(t) grows exponentially by comparison it expands super-luminally. If vacuum energy dominates from an initial time defined to be $t_i = 0$ to $t_f$, then the physical horizon distance (Table2) is approximately $(c/H)[e^{Ht_f} - 1]$. There is an exponential factor compared to $D_H$ for matter or radiation domination. The final conformal time during this period, of duration $t_f$, and constant H is approximately;

$$\tau_f \sim ce^{Ht_f}/[a(t_f)H] \tag{10}$$

The value of conformal time during a very limited period of acceleration can be comparable to the expected value of approximately 3 for the entire later evolution of the Universe, Fig.1. This fact allows for the solution of causality issues in BB cosmology by invoking a limited period of inflation.

## 5. Radiation and Matter Domination and the CMB

Given the dominance of matter over radiation at the present time (Table1), the scaling of a(t) with time can be used to project backward. The radiation term will become more important and will, at some past time, become dominant. Therefore, the Universe is expected to be very hot and dense at early times which implies the Big Bang. The value of the scale factor a(t) can often be normalized to the present value, $a_o$ making the overall scale irrelevant. For Fig.1 and Fig.2 a simple matter or radiation domination is assumed to hold making for the simplest power law extrapolations back in time.

Numerical results are obtained by assuming the present time, $t_o$, is 13.81 billion years since the Big Bang. The present energy densities are as defined in Table1. Assuming matter domination, the inverse matter density goes as t squared while the inverse radiation density rises as t to the 8/3 power. Therefore, at earlier times radiation will dominate the energy density. The two densities become equal at an earlier time, $t_{eq}$, when the energy density is 5.6 x $10^{11}$ GeV/$m^3$ and the temperature of the radiation, scaling inversely as a(t), is 18,600 degrees Kelvin. That temperature corresponds to an energy of 1.5 eV. The time $t_{eq}$ is 8 x $10^{11}$ sec or 24,000 years. It is assumed that the "dark matter" which now predominates over ordinary matter has the same a(t) behavior as ordinary matter. Before $t_{eq}$ radiation is the dominant form of energy in the Universe. Dark energy is ignored in this simple approximation.

One can distinguish between $t_{eq}$ and the decoupling or recombination time when the plasma of photons and electrons, which is opaque to light, becomes a system of non-interacting hydrogen atoms and photons, transparent to light. The decoupling time occurs when the plasma scattering width falls behind the expansion, $\Gamma_\gamma < H$, ending thermal equilibrium.

Because the photon to baryon ratio is so large and because there is a long tail to the Maxwell-Boltzmann energy distribution for the photons, the CMB appears later than $t_{eq}$, at a mean photon energy of about 0.32 eV or a time $t_{dec}$ about 327 thousand years or about $10^{13}$ sec. A more complete analysis leads to a value for $t_{dec}$ of 1.2 x $10^{13}$ sec [14]. The light from this epoch is the CMB.

Recombination occurs when the electrons and baryons combine into neutral atoms. The recombination, decoupling and last scattering of the photons all happen at about the same time and $t_{dec}$ is used as a shorthand for all three epochs when the scale a(t) is about 1100 times smaller than at present. The physical horizon at the time of formation of the CMB was therefore about 1100 times smaller than at present and subtends an angular region on the sky of about one degree. The opacity of the baryon-photon fluid before decoupling means that the CMB is the earliest object available for study barring a possible future ability to detect gravitational waves or primordial neutrinos which have no electromagnetic interactions and are therefore transparent for vey early times.

The time can be scaled to periods earlier than $t_{eq}$ reliably since the atomic and nuclear physics which is in play is well understood. Going to 0.01 sec, the energy density increases to 2.5 x $10^{39}$ GeV/m$^3$ and the radiation temperature is 1.5 x $10^{11}$ degrees Kelvin or 13 MeV. This energy scale remains in the domain of nuclear physics and should be reliable. The energy density of matter and radiation from 0.01 sec to the present is shown in Fig.2. Also shown are representative densities of water, the sun and a white dwarf star. The energy of the radiation component of the Universe is shown in Fig.3. It goes as the inverse of the scale a(t), but the time dependence of a(t) depends on whether the Universe is radiation or matter dominated. Two representative scales are also shown; the atomic scale is the ionization energy of hydrogen while the nuclear scale is set by the binding energy of the deuteron.

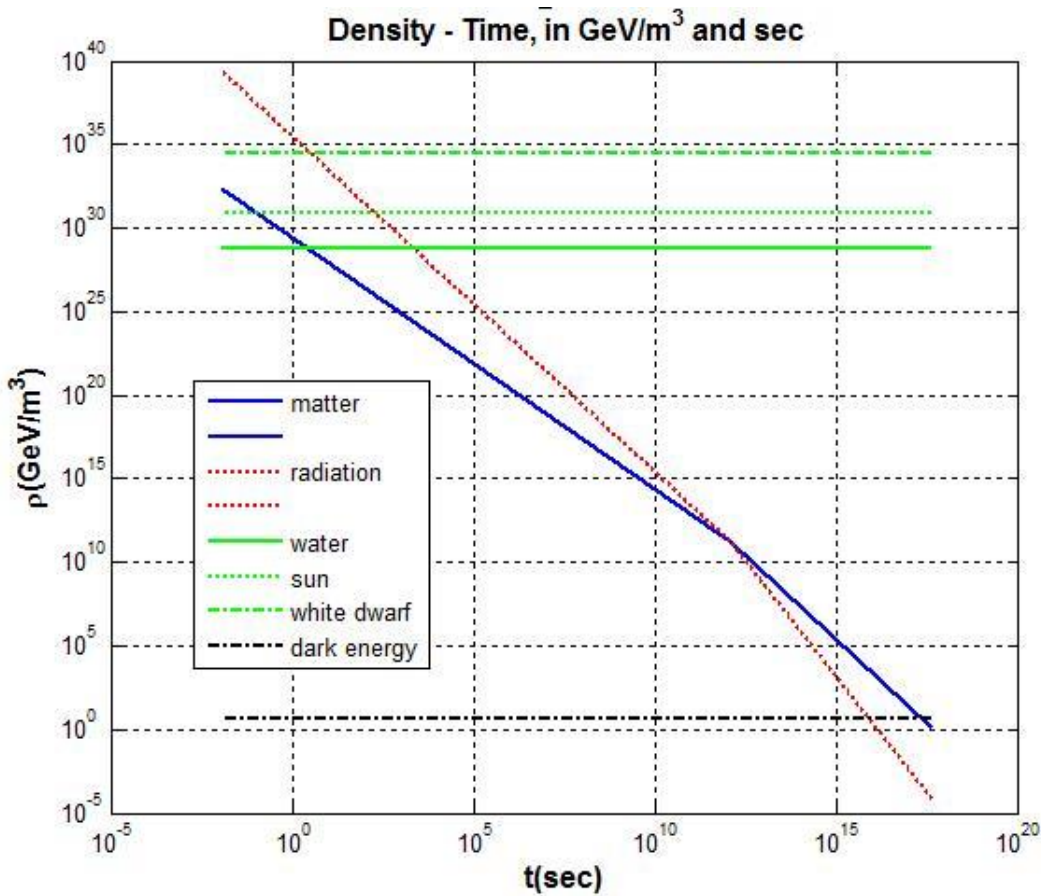

Figure 2: Time dependence of the energy density for matter, radiation and dark energy. The radiation and matter densities are equal at a time $t_{eq}$ = 0.8 x $10^{12}$ sec. Horizontal densities of dark energy, water, the sun and a white dwarf are shown to set the density scale.

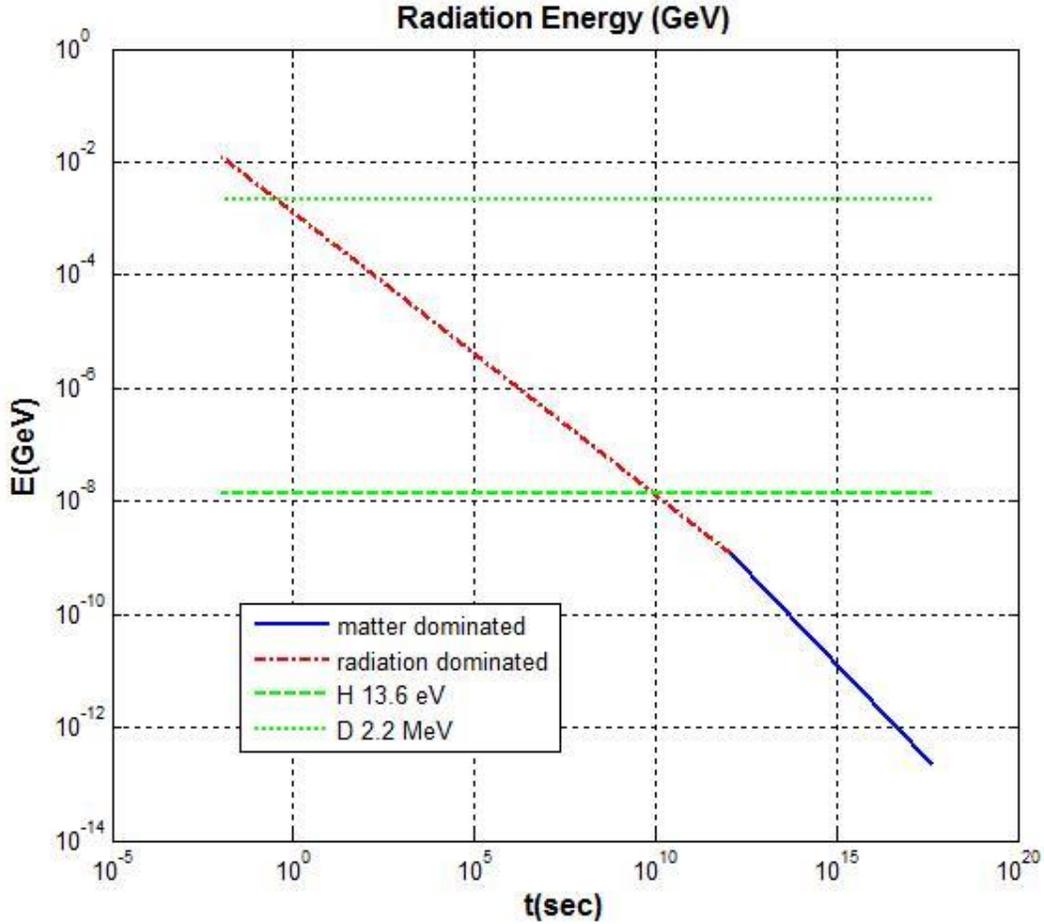

Figure 3: Time dependence of the energy (Temperature) of the radiation. The horizontal scales for atomic and nuclear physics are shown to set the scales.

## 6. The Flatness and Causality Issues for BB Cosmology

There are two major issues for Big Bang (BB) cosmology [15]. One is called the flatness problem. If the Universe is not spatially flat, as was previously assumed in conformance with the present data, the radial term in the Robertson-Walker metric is modified by a parameter k to be $dr^2/(1 - \kappa r^2)$, Eq.2. The κ parameter defines the curvature which may be -1, 0 or 1. The current value of the flatness parameter [14] is that the Universe is within a few percent of the critical energy density which divides positive and negative curvature so that κ is near zero.

The problem is that if the curvature is now near to flatness, then in the past it must have been incredibly fine-tuned to be almost exactly zero. Eq.3 defines the critical value of ρ because κ equal to zero was assumed there. The value for κ more generally is defined in terms of the present value of the Hubble parameter H, the energy density, and the Planck mass. The relationship of H to density has an added term proportional to κ.

$$(\dot{a}/a)^2 = (8\pi/3M_p^2)\rho - \kappa/a^2 = H^2$$
$$\rho_c = H_o^2/(3M_p^2/8\pi) \qquad (11)$$
$$\Omega = \rho/\rho_c, \ |\Omega - 1| = \kappa/(Ha)^2 = \kappa d_H^2$$

The deviation from flatness is controlled by the square of the value of Ha. As seen from Table2, H goes as n/t while a(t) goes as $t^n$, so that Ha goes as $nt^{n-1}$. For n less than 1, which holds for radiation or matter dominance, Ha decreases as t increases so that the curvature parameter, $|\Omega-1|/\kappa$ increases at late times. Indeed, Ha for matter domination goes as the inverse 1/3 power of t while for radiation domination it goes as the inverse square root of t. Therefore a flat space is unstable as time increases and fine tuning seems difficult to avoid as a necessary initial condition on the Universe. Note that, from Table2, vacuum energy behaves oppositely driving the flatness parameter to zero which indicates a possible solution to the flatness problem.

There is only one parameter, a(t) in the flat metric we have assumed. In what follows, a more exact numerical integration of the equations defining a(t) is made, going beyond the first approximation which used simple radiation or matter domination. An intermediate approach, appropriate near the present time, could be used since an analytic solution for H(t) can be found if only matter and vacuum energy are active which generalizes the simple power law behavior [6] which will not be pursued. The full evolution equation for a(t) to solve follows from Eq.3 and the behavior of a(t) shown in Table2, and is;

$$\alpha = a(t)/a_o$$
$$\dot{\alpha} = d\sqrt{1/\alpha + b/\alpha^2 + f\alpha^2} \qquad (12)$$

The constant is, $d = \sqrt{(8\pi/3)M_p^2 \Omega_m \rho_c}$ = 1.18 x $10^{-18}$ /sec. The other terms are b equal to the ratio of radiation to matter density at present and f equal to the ratio of vacuum to matter density at present. The result of this numerical integration of the effects of all three sources of energy treated simultaneously appears in Fig.4 which shows the evolution of a(t)/$a_o$ with time from 1 sec to a time slightly later than the present where the effects of dark energy become manifest. The plot shows the ratio of a(t) to $a_o$. The present time and $t_{eq}$ are shown as vertical lines. Note that at future times the vacuum energy begins to drive an exponential expansion of a(t).

The flatness parameter is defined by the behavior of $(Ha)^2$ as seen in Eq.11. For vacuum energy domination H is a constant and the scale factor is exponentially increasing, as seen in Eq.6. Therefore a vacuum energy density would drive the $1/(Ha)^2$ rapidly to zero. This behavior is seen in Fig.5 at future times when vacuum energy fully dominates. In the past, the curvature, $|\Omega-1|/\kappa$, increased by a factor roughly $10^{15}$ from a time of one second to the present. The expected linear dependence of the flatness parameter on t, $t^{2(1-n)}$, is seen in Fig.4 for early times, dominated by radiation with a t dependence, followed by a softer $t^{2/3}$ behavior for matter dominated times.

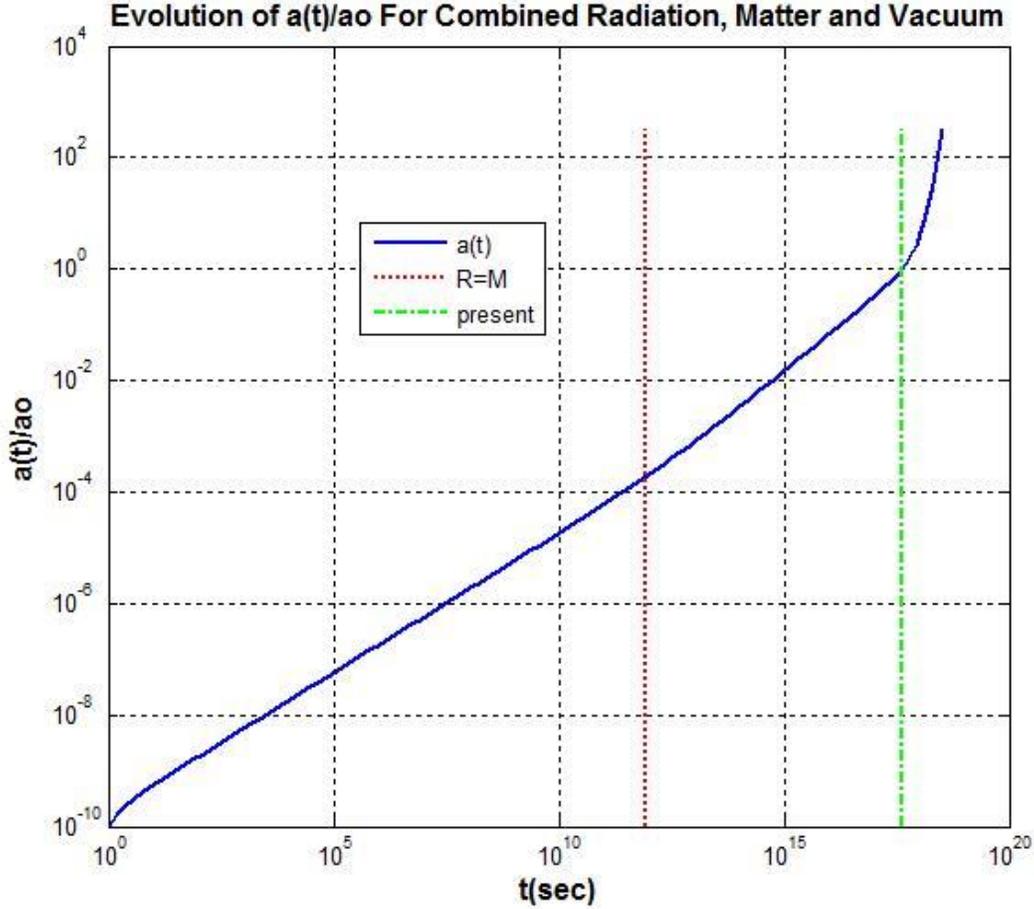

Figure 4: Time dependence of the scale factor $a(t)/a_o$ from 1 sec to the present and a bit later. All sources of energy density are treated simultaneously. The vertical lines indicate $t_{eq}$ (dotted) and $t_o$ (dash-dot).

An estimate of the inflationary expansion needed to solve the curvature problem can be made assuming a constant value of H, as occurs for a cosmological constant, operating for a short period of time, $t_f$, early in the history of the Universe. The curvature term at $t = 0$ is assumed to be of order 1, $|\Omega-1|\sim 1$. The quantity $\kappa = |\Omega-1|(aH)^2 = |\Omega-1|/(d_H)^2$ is a constant, Eq.11. Assume that $N = 60 = Ht_f$ which will be motivated later. The scale increases by a factor $e^N \sim 10^{26}$ during inflation which means that $|\Omega-1|\sim 10^{-52}$ after inflation ends.

This is effectively the initial condition for BB cosmology. After this inflationary period assume radiation domination until $t_{eq}$ at $\sim 10^{12}$ sec. Since $(Ha)^2$ goes as t in this period $|\Omega_{eq}-1|\sim 10^{-40}/t_f$. From $t_{eq}$ until the present at $t_o$ the scaling goes as $t^{2/3}$ leading to a present value of $|\Omega_o-1| \sim 5.8 \times 10^{-37}/t_f$. Assuming that the present experimental deviation from flatness is only 1%, the scale for the time of the end of inflation is $t_f \sim 5.8 \times 10^{-35}$ sec in order to just solve the flatness problem.

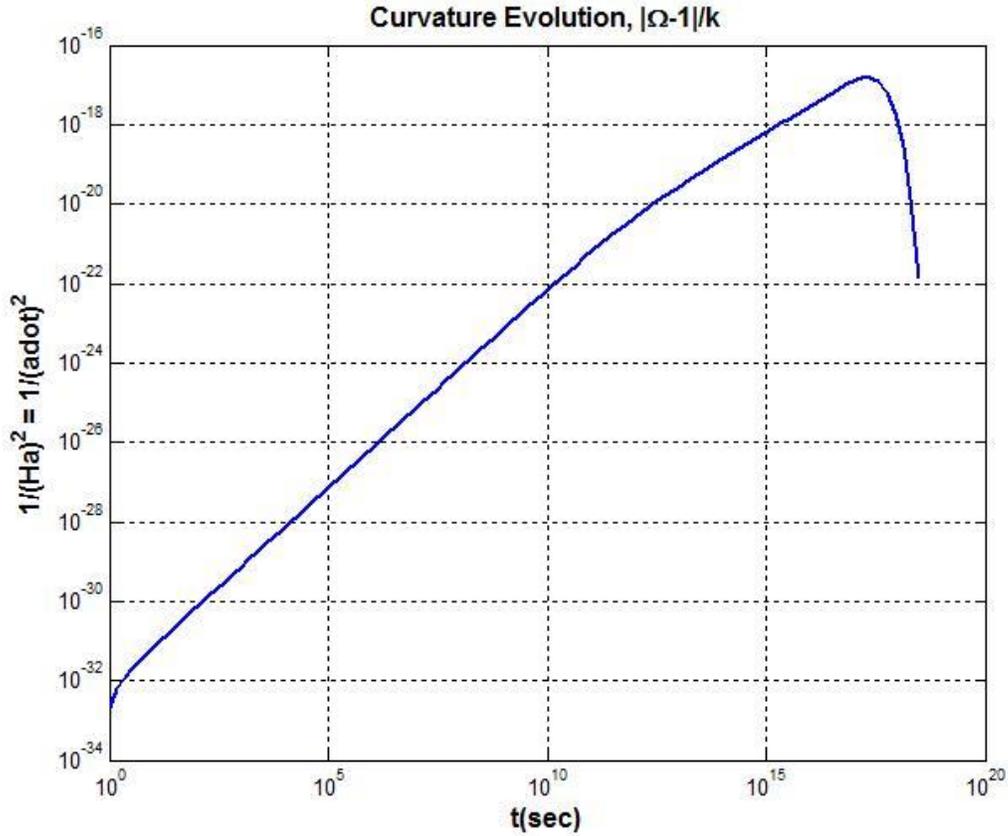

Figure 5: Evolution of the curvature parameter from a time of 1 sec to the present and near future. Note that vacuum energy drives down the curvature which grows rapidly in the matter or radiation dominated epochs. The absolute vertical scale is arbitrary.

The causality issue in BB cosmology has to do with the observed uniformity of the CMB. A rough order of magnitude estimate is as follows. Consider the CMB and extrapolate back to the decoupling time, $t_{dec}\sim$ ten times $t_{eq}$, when the scale was about $10^{-3}$ times the present scale (Fig.4). The Hubble distance $D_H$ at present, Table1, corresponds to 1.37 x $10^{23}$ m at the time of CMB decoupling. However light has only gone, since the assumed t = 0 Big Bang, a distance of about $ct_{dec}$ or 3 x $10^{21}$ m. There is a mismatch by about a factor of 45. Hence, there is no way that the CMB can be causally connected which is needed to easily explain its' temperature uniformity. That conclusion assumes that no new physics intervenes going back to time zero or the "Big Bang".

The dimensionless conformal time defines the horizon and it displayed from one second to the present in Fig.5. This time dependence was derived using the numerical results of Fig.4 assuming a $t_o$ value quoted above. The expected behavior that conformal time goes as the square root of t at early times and the cube root at later times, Table2, is observed. Numerically, the conformal time from one sec to $t_{dec}$, the decoupling time is 0.062, while from $t_{dec}$ to $t_o$ it is 2.96.

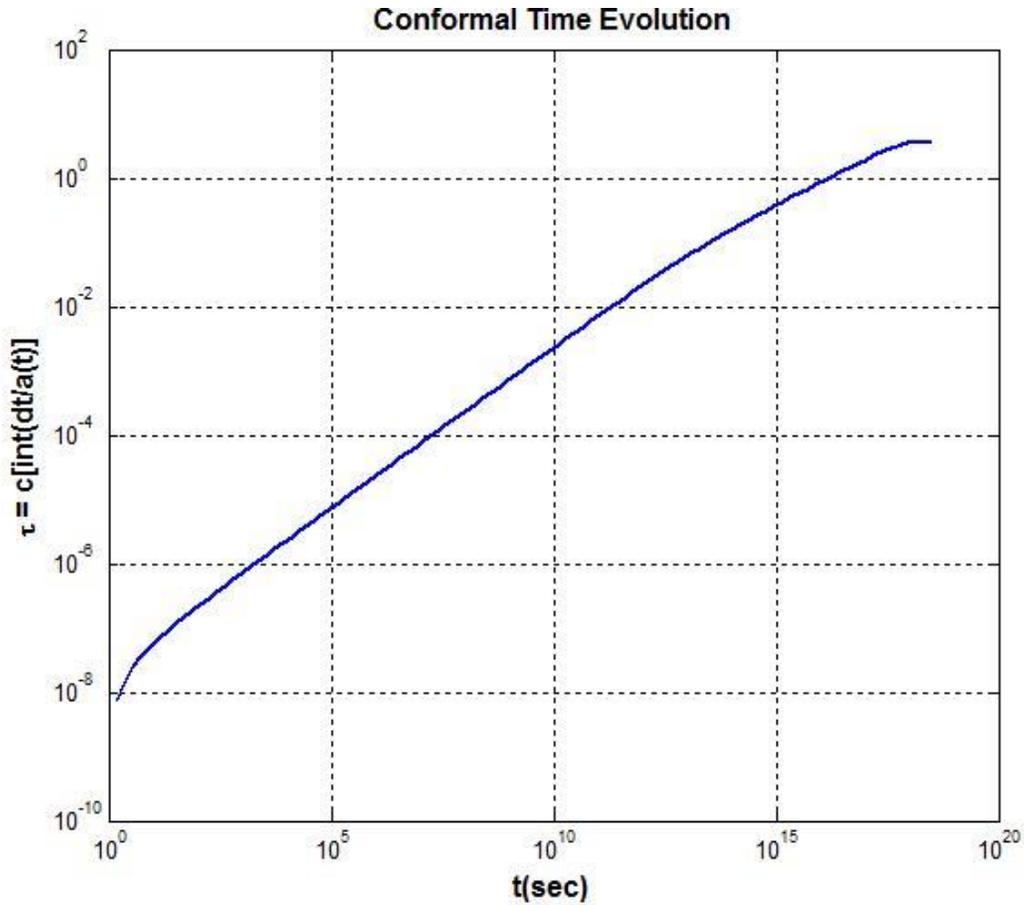

Figure 6: The evolution of conformal time from t = 1 sec to the present time and somewhat beyond, t = $10^{18.5}$ sec, as derived from the numerical results of integrating Eq. 3 with all three sources of energy, matter, radiation and vacuum.

The conformal time now is about 3.02. The initial conformal time is defined to be zero. Light travels on straight lines inclined by 45 degrees (the light cones), Eq.8, in the (τ, r) plane. Causally connected events occur within the past light cones as in the case of special relativity (SR). The cone for the present is shown in Fig.7. Also indicated is the conformal time at ~$t_{dec}$ or decoupling time, of 0.062 for the CMB conformal time about 48 times less than the present time (Fig.6). Clearly, events spanning the CMB range of r extrapolated to zero conformal time cannot be causally connected since the CMB conformal time light cones do not reach the (0,0) origin. The mismatch factor approximately agrees with the previous rough estimate.

One advantage of a more exact numerical treatment of the time evolution of the scale a(t) is that the behavior of H(t)t becomes more transparent. The results are shown in Fig.8, where the expected approximate constancy with t, H(t)t = n, is seen for radiation and matter domination. The rise for t near $t_o$ is due to the dark energy contribution which dominates at the present time.

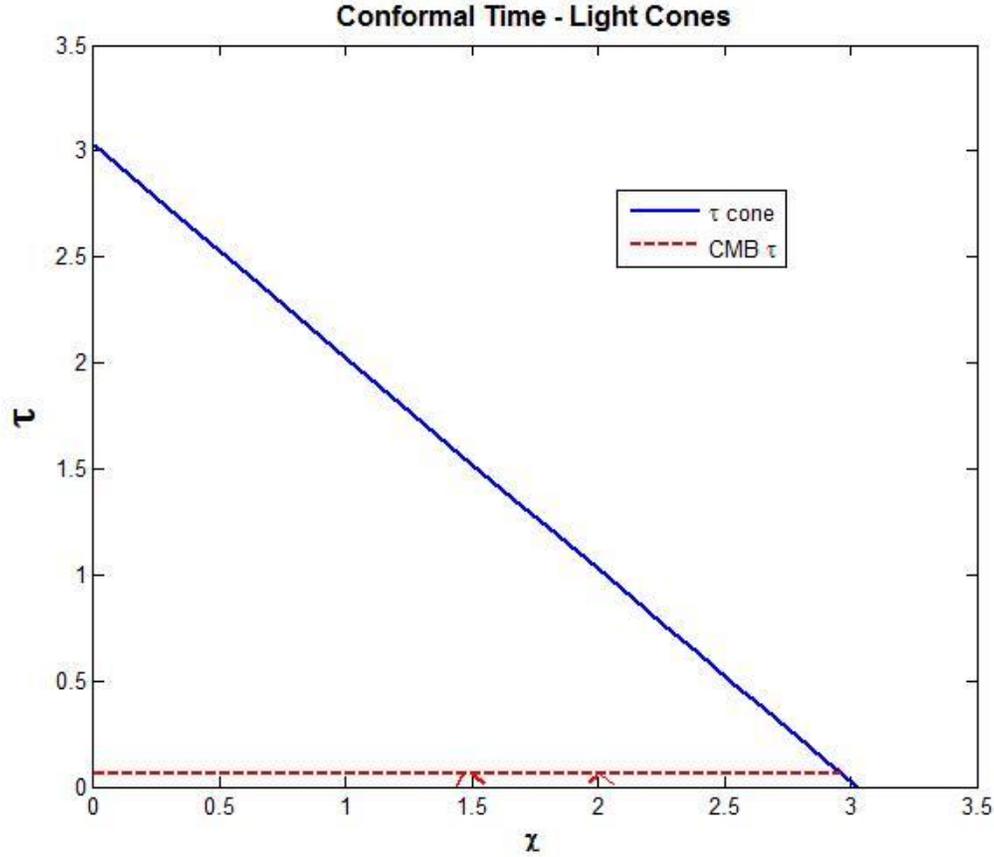

Figure 7: Light ray conformal time from the present time, $\tau = 3.02$, back to the CMB, $\tau = 0.062$. The CMB conformal time backward light cones are shown for a few points. They cannot communicate with the (0,0) origin or with each other except very locally.

This numerical result confirms the analysis which determined $t_o$ to be fortuitously very close to the value of $t_H$.

The causal problem can be solved if there is a limited inflationary period when a small, causally connected, patch of space is inflated to become the presently observable Universe. A rough estimate of the inflation parameters can be made using Eq.10 and the results of a similar exploration of the flatness problem which, by assuming $N = Ht_f = 60$, estimated $t_f = 5.8 \times 10^{-35}$ sec or $H = 1.03 \times 10^{36}$ sec$^{-1}$ are needed to solve the flatness problem. The scale needed now to have a conformal time growth during inflation of 3.02 in order to just solve the causality problem is then, $a(t_f) = ce^N/\tau_o H$ or $a(t_f) = 0.011$ m or an initial scale factor of $1.1 \times 10^{-28}$ m. These estimates for $t_f$, H and $a(t)$ set the approximate scales for a more realistic model for inflation which is explored later.

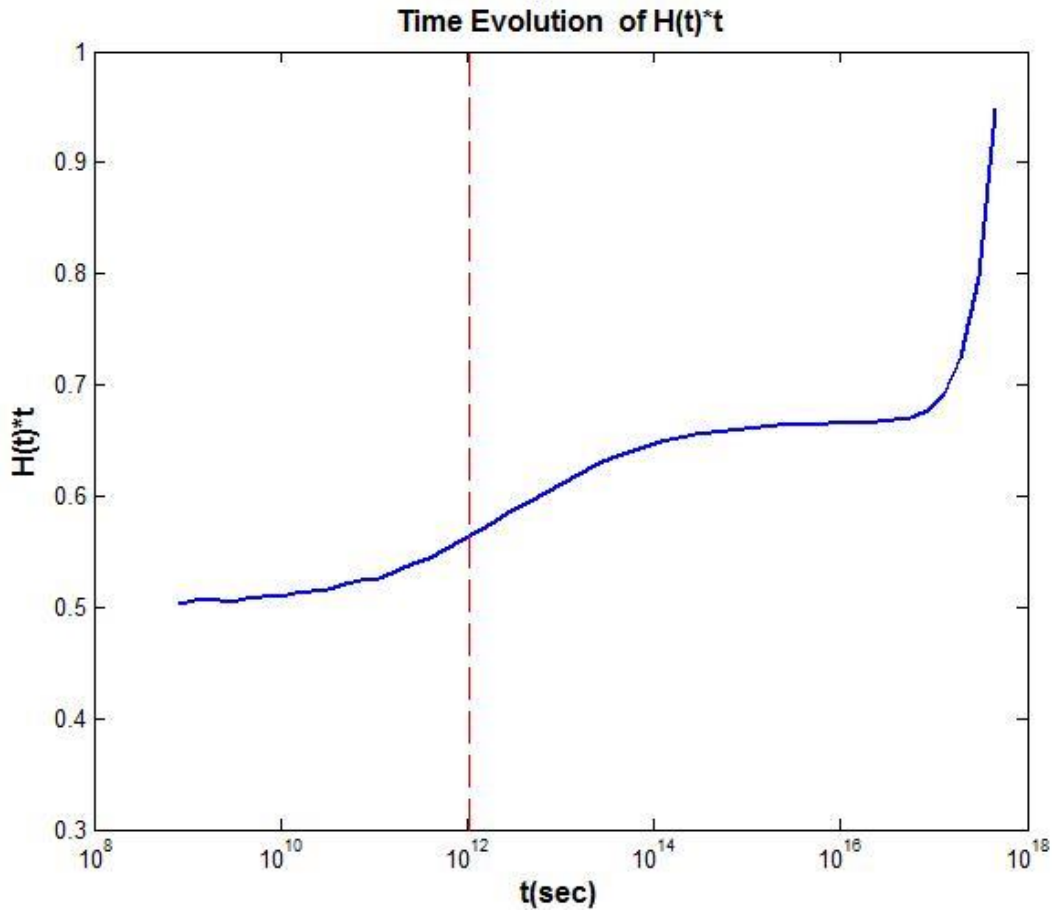

Figure 8: Plot of the behavior of H(t)t as a function of time. During radiation domination, Ht is ~ ½, while for matter domination Ht ~ 2/3. At late times the domination of the dark energy causes a rise in the value of Ht until $H_o t_o$ ~ 1.0 at present. The dashed red vertical line displays $t_{eq}$.

## 7. Inflation and Scalar Fields

The solutions to the problems of BB cosmology using a cosmological constant existing for a limited time were previously estimated. Inflation aims to solve the BB problems by postulating a rapid expansion of the scale factor at very early times, thus invalidating the simple extrapolation to time zero assumed in BB cosmology. New physics intervenes at very early times. As will be discussed, inflation is the ultimate free lunch where the initial potential energy of a scalar field with quantum fluctuations becomes the richly structured observed Universe.

In the approximation of a cosmological constant, $\rho_v$, H is a constant for all time. Therefore a dynamical mechanism for the limited time of inflation is needed. The physical mechanism for the existence of an approximately constant value of H which lasts for a limited time posits a scalar field which has a large initial potential energy which depends on the potential shape such that it is sensibly constant (called "slow roll") for a sufficient period of time to solve the flatness and causal problems before it speeds up and falls into the minimum of the potential. Then the field oscillates about its' minimum decaying into less massive particles insuring that $t_f$ is finite. The physics of the large initial potential, the shape of the potential and the evolution of the field are unknown and are here simply assumed ad hoc.

This postulated field is not definitively identified with a well-established field known to high energy physics. However, the Higgs field is a fundamental scalar field and is a possible candidate under the assumption of non-minimal coupling of it to gravity as will be discussed later. This is a very interesting, compact and economical possibility that low energy Higgs physics is an agent of Planck energy scale inflation physics.

The field, φ, is approximately uniform in space, but decreases with time [16]. The present Universe, remaining after inflation, was a small homogeneous initial patch which was causally connected and inflates to become the homogeneous CMB. The dimension of the field is mass as can be inferred from noting that it appears as a kinetic term, $(d\varphi/dt)^2$, in the Lagrange density whose space and time integral leads to a dimensionless action. Since this kinetic term is therefore an energy density, it has the dimension of mass to the fourth power. The kinetic term is a special case of the Klein-Gordon kinetic energy, $\varphi^* \partial_\mu \partial^\mu \varphi$, for a spatially uniform field.

The field energy density and pressure are; $\rho_\varphi = \dot{\varphi}^2/2 + V(\varphi), p_\varphi = \dot{\varphi}^2/2 - V(\varphi)$ where V(φ) is the potential energy of the scalar field. Allowing for a pressure source term, Eq.5, the acceleration of a(t) is positive and inflation occurs if the potential term dominates. The Hubble parameter, density time derivative, and scale factor acceleration are special cases of Eqs.3, 4 and 5;

$$H^2 = (8\pi/3M_p^2)(\dot{\varphi}^2/2 + V(\varphi))$$
$$\dot{\rho} + 3H\dot{\varphi}^2 = 0$$
$$\ddot{a}/a = -(8\pi/3M_p^2)(\dot{\varphi}^2/2 - V(\varphi)) \qquad (13)$$

If the potential dominates and varies slowly with time, H is quasi-constant and the acceleration is positive. Using Eq.13 and differentiating with respect to time, the equation of motion for the field is;

$$\ddot{\varphi} + 3H\dot{\varphi} + dV/d\varphi = 0 \qquad (14)$$

The field acts like a harmonic oscillator, with H supplying the damping or frictional effect since expansion makes smooth any oscillation. The restoring force is supplied by the derivative of the potential with respect to the field.

The length of time for inflation to be active depends on the shape of the potential, dV/dφ. Neglecting $\dot{\varphi}$ in the expression for H and the second time derivative in the harmonic equation is called the slow roll approximation, $H^2 \sim (8\pi/3M_p^2)V(\varphi), 3H\dot{\varphi} \sim -dV/d\varphi$ which is assumed to hold initially. Dimensionless parameters which characterize the first and second derivative of the scalar potential approximately specify the shape of the potential and are;

$$\varepsilon = (M_p^2/16\pi)[(dV/d\varphi)/V]^2$$
$$\eta = (M_p^2/8\pi)[(d^2V/d^2\varphi)/V] \quad (15)$$

The shape parameters are related to a(t), H(t) and the acceleration of a(t) which shows that inflation ends when ε = 1; $\varepsilon = -\dot{H}/H^2, \eta = -\ddot{\varphi}/H\dot{\varphi}$ and $\ddot{a}/a = H^2(1-\varepsilon)$.

In order to solve the BB flatness and causality issues a sufficient expansion of the scale factor, a(t), is needed. It is characterized by the number of "e-folds", N. Using Eq.13 in the slow roll approximation, $\int H dt \sim \int 3H^2 d\varphi/(dV/d\varphi)$;

$$a \sim e^{Ht}, N \sim \int H dt$$
$$N = (8\pi/M_p^2)\int [V/(dV/d\varphi)]d\varphi$$
$$= (2\sqrt{\pi}/M_p)\int d\varphi/\sqrt{\varepsilon} \quad (16)$$

The number of "e-folds" depends on how fast the field is fractionally decreasing. The number of e-folds, N, is inversely proportional to the square root of the slow roll parameter ε or proportional to the inverse fractional change of the potential with the field, V/(dV/dφ). The field is active only at very early times and then disappears so that the successes of BB cosmology are retained. The exponential increase in a(t) drastically reduces the temperature since Ta is a constant. After the field disappears, the Universe will need to re-heat to the high temperatures needed to create the light nuclei whose relative abundance is predicted by BB cosmology and is a major success of the BB model.

## 8. Quadratic Potential

A simple model for the potential field has a quadratic dependence on the field similar to the case for an harmonic oscillator, $V(\varphi) = m^2\varphi^2/2$. The unknown field mass is m. This model is simple and in accord with present cosmological data. The slow roll parameters for this potential depend inversely on the square of the field, $\varepsilon = \eta = (1/4\pi)(M_p/\varphi)^2$. To achieve a sufficiently large value of N, an initial field larger than the Planck mass is needed. Using Eq.13 in the slow roll approximation, the time dependence of the field can easily be solved for;

$$\dot{\varphi} = -mM_p / \sqrt{12\pi}$$
$$\varphi = \varphi_i - (mM_p / \sqrt{12\pi})t \quad (17)$$

The characteristic time for the field to exist is set by m since m defines the time rate of decrease of the field. Very approximately the field "decays" in a time $\sim \sqrt{6N}/m$. The subscripts i and f refer to the start and end of the inflationary period, $t = t_f - t_i$. Neglecting the final field remaining when the field variation is no longer a slow roll, the number of e folds is defined by the size of the initial field, Eq.16, $N = 2\pi(\varphi_i / M_p)^2$. The initial field value in this model will be larger than $M_p$ to achieve ~ N ~ 60. The Hubble parameter depends linearly on the field and the field mass while the scale parameter depends exponentially on the field mass and the time integral of the field;

$$H = \sqrt{4\pi/3}(m/M_p)\varphi$$
$$a(t) = a_i e^{\sqrt{4\pi/3}(m/M_p)\int \varphi dt} \quad (18)$$

The curvature discussion for constant values of H indicated that a time of inflation of about $10^{-34}$ sec was required with an H of about $10^{36}$ $sec^{-1}$ and with a final scale factor when inflation ends of about 0.01 m. The quadratic scalar field scenario is only in very approximate agreement with these estimates because it has a field which decreases linearly in time so that, Eq.18, H also decreases linearly during inflation.

## 9. Quadratic Potential and Inflation

At present, fundamental particle physics is understood up to about the one TeV scale. Scaling from $t_{eq}$ of about $10^{12}$ sec, with 1.2 eV energy, scaling in energy to one TeV decreases a(t) by about a factor of $10^{12}$ and t decreases by a factor of $10^{24}$ going down to about $10^{-12}$ sec.

The postulated physics of inflation operates at a vastly increased scale of energy and a vastly shorter time period, estimated to be at $10^{-34}$ sec, quite beyond what is now understood. Much new physics could intervene, such as superstrings, supersymmetry, and grand unification. In this note a radiation or relativistic particle dominated regime is assumed to apply immediately after inflation ends until matter begins to dominate at $t_{eq}$. This assumption is clearly very simplistic.

Numerical estimates for inflation are made only in the quadratic field model. Later other models are explored, but this model agrees with present data and therefore serves as a plausible example. With only a few parameters, the initial field value and the mass, this model agrees with present data on CMB scalar power, CMB spectral index, CMB tensor power, N, and CMB oscillations.

Inflation ends when the acceleration stops, when ε is one or $\varphi_f / M_p = 1/\sqrt{4\pi}$ = 0.28. A sufficiently large value of N is needed to solve the BB issues, N ~ 60, which sets the scale, Eq.16, for the initial field, $\varphi_i / M_p = \sqrt{N/2\pi}$ = 3.1. A value of 3.25 was chosen for the numerical estimates used here. The value of the H parameter as a function of time depends on the scalar mass, m, since it defines the time dependence of the field, Eq.17, and thus the scale factor a(t), Eq.18. A value of $m/M_p$ of $10^{-6}$ is chosen. That choice leads to a slope of the field of 2.96 x $10^{36}$ sec$^{-1}$ and a time when inflation is active of about $\varphi_i$/slope = 1.10 x $10^{-36}$ sec.

The evolution in the inflationary regime is tracked from $10^{-38}$ to $10^{-36}$ sec. Other choices of the range of time can be made for other choices of m. However, the model must explain the CMB temperature fluctuations which define a value for m as will be shown later. The Hubble parameter is initially about 6.6 x $10^{-6}/M_p$ and decreases linearly to about 0.6 x $10^{-6}/M_p$ at $10^{-36}$ sec. This decrease gives the model a definite end to the period of inflation.

The values of a(t) as a function of time were determined by a numerically and are shown in Fig.9. An approximate exponential increase appears as is expected from the simpler case with a constant value of H. The value of a($t_i$) is set arbitrarily to be (c$t_i$)/4 or 7.5 x $10^{-31}$ m. The increase in a(t) by a factor of 1.2 x $10^{28}$ over the inflationary period supplies 64.6 e folds. At the end of inflation the scale is a($t_f$) = 0.009 m.

The curvature as a function of time, Eq.11, is shown in Fig.10. During the inflationary period the increase of a(t) by a factor 1.2x $10^{28}$ and the approximate constancy of H mean that the curvature decreased by a factor of 1/ 8.9 x $10^{-55}$. After the scalar field has "decayed" into Standard Model (SM) particles, a radiation dominated curvature was assumed where $(1/Ha)^2$ scales as t (Table2). In contrast, for the BB cosmology the flatness, |Ω-1|/k would have to be fine- tuned to be less than $10^{-64}$ at the Planck time. At the starting time of inflation the curvature is unknown and a plausible value of |Ω-1| = 1 was chosen. The value extrapolated to the present is 0.005, comparable to the current limit [14], so the curvature problem is explicitly solved for this example. Inflation resets the initial unknown value of the curvature to be a value about $10^{54}$ times smaller at the end of inflation. After that a linear increase with t is assumed until $t_{eq}$ followed by a n = 2/3 power law for matter, leading to a value 0.005 at present.

$D_H$ defines the radius of the causal event horizon. Because of decelerated expansion, the power law behavior of a(t) leads to a Hubble length of ct/n for a(t) ~ $t^n$. In a period of inflation, the Hubble distance is approximately constant (Table2). In the quadratic field model the Hubble distance is displayed in Fig.11 where an abrupt transition from inflation to radiation dominance is assumed to occur because the field is assumed to decay rapidly. During inflation a physical scale, in this case a(t)/1000 grows rapidly and becomes larger than the Hubble distance $D_H$ which is roughly constant during inflation. After inflation the Hubble distance grows as ct/n or 2ct in a radiation dominated regime while a(t) grows more slowly, as the square root of t. During the time that any physical scale is greater than the Hubble distance that scale is outside of causal influence. As seen in Fig.11 at a later time a physical scale becomes less than the Hubble distance, re-enters the horizon and then can become causally active again. For this particular scale, the crossover occurs at about $10^{11}$ sec, somewhat before $t_{dec}$ of $10^{13}$ sec when the CMB is formed. The Hubble distance has a delayed increase with time since growth of c/H occurs only after inflation.

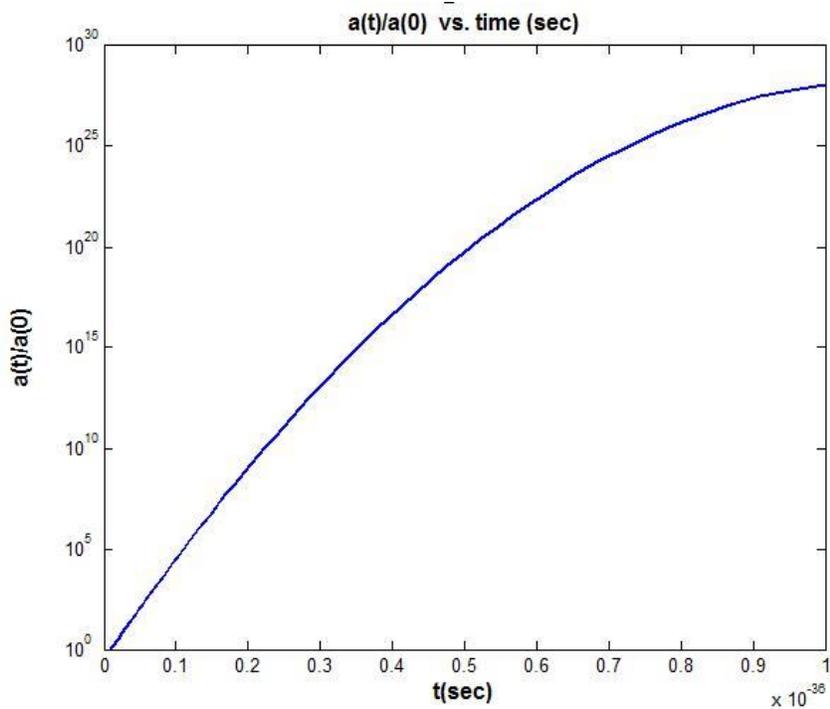

Figure 9: Time dependence of the scale factor a(t) during the inflationary period, where a(t$_i$) is here taken to be 1. An approximate exponential behavior is evident.

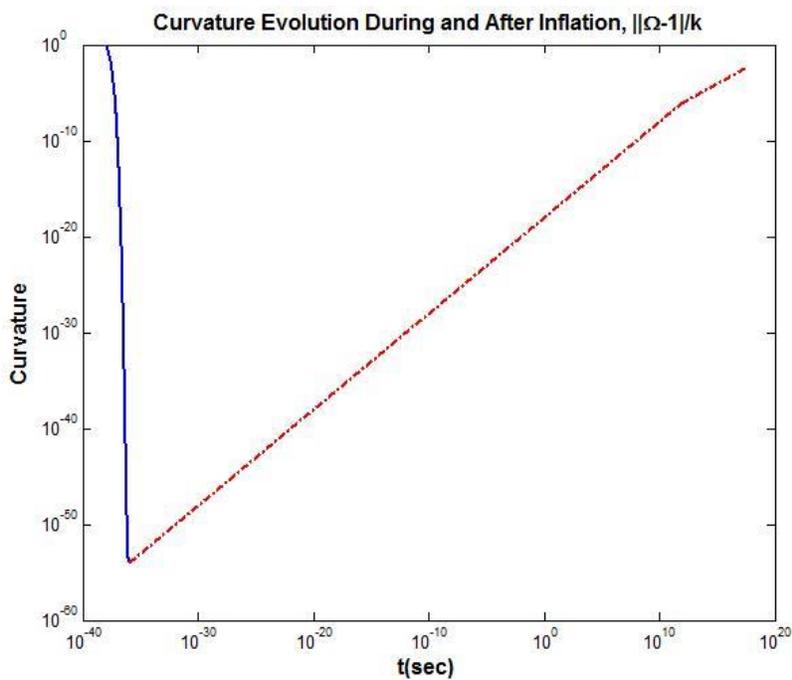

Figure 10: Time dependence of the spatial curvature during the inflationary period followed by an assumed radiation dominated period

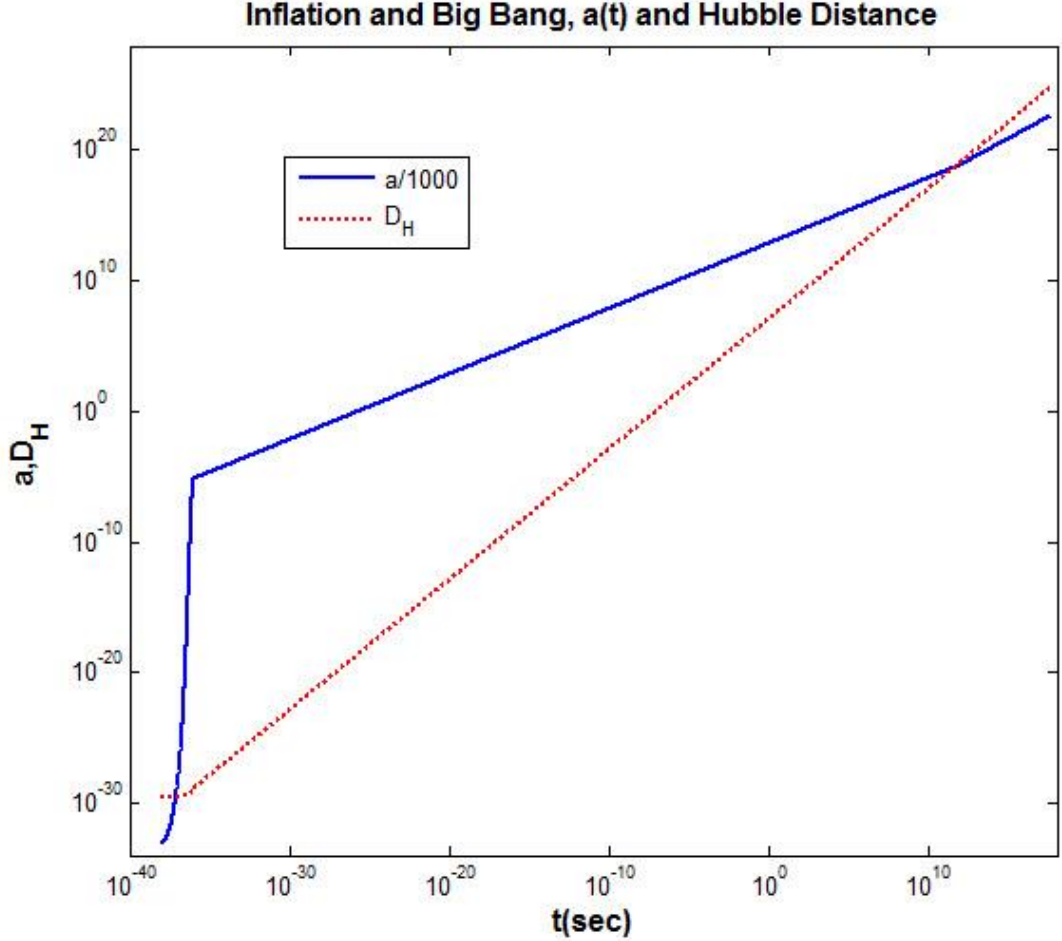

Figure 11: Time dependence of the physical scale factor a(t)/1000, solid blue, and the Hubble length, $D_H$ (m), dotted red, for an inflationary period followed by a period of radiation dominance.

At the end of inflation in the present numerical exercise $c/H(t_f)$ is 1.5 x $10^{-29}$ m. At $t_{eq}$ it is 1.4 x $10^{19}$ m, while at present it is scaled to $c/H_o$ of 6.9 x $10^{24}$ m. In contrast, $a(t_f)$ is 0.009 m. At $t_{eq}$ a(t) is 8.2 x $10^{21}$ m, while at the present it is 5.2 x $10^{25}$ m. The estimated present value of $a_o$ is only a factor 2.5 less than the assumed value of $a_o$ (Table1). The extrapolated value for $D_H$ is a factor 20 lower than the measured present value. That discrepancy arises from the choices made for the parameters of the model, $\phi_i, m, t_f, a(t_i)$. For example, increasing the value of m, Eq.18, would increase the value of H during inflation and therefore raise the estimate of the present value. The aim of the exercise is not a fit to the data but a plausibility study for a specific and simple model.

With the resetting of clocks to time zero before inflation, the conformal time during inflation covers about 3.20 units in the present example. The time development of the conformal time during inflation is shown in Fig.12. As expected from the estimates for conformal time in the constant H model, the present numerical estimate shows that causal connection over the whole Universe at the time of the CMB is possible with N ~ 60 e-folds thus explicitly showing that the present simple model solves the BB causality issue.

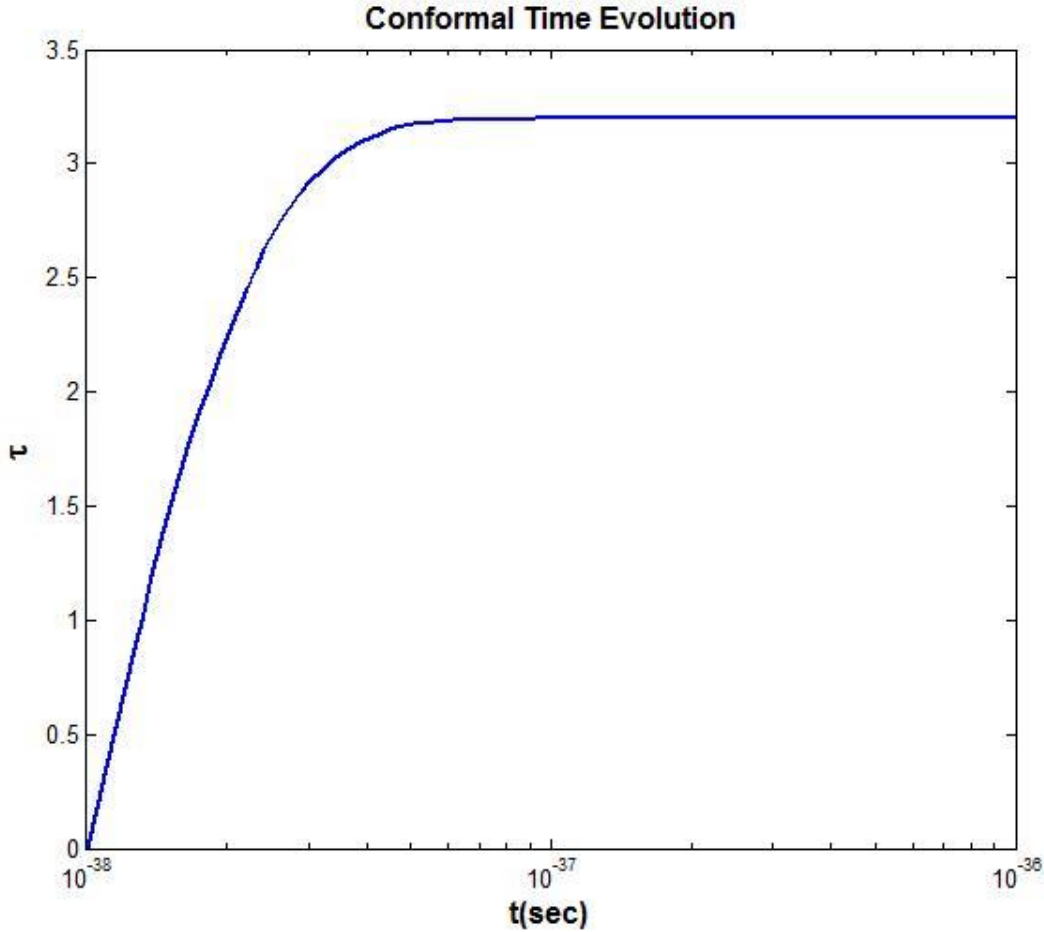

Figure 12: Conformal time (dimensionless) as a function of coordinate time during inflation. There is a sufficient range in conformal time, > 3.0, to allow CMB causal communication, Fig.7, at very early times.

The causal issues can be examined using either comoving or physical quantities, horizons $D_H$ or $d_H$. It is perhaps easier to visualize the situation using comoving quantities such as conformal time. The conformal Hubble horizon, $d_H$, which is dimensionless, decreases dramatically during inflation, since H is roughly constant and a(t) is growing exponentially. At $t_f$ it has a value in this example of 6.6 x $10^{-27}$. Scaling to $t_{eq}$ the value is 0.006, while at present it is estimated to be 0.486. The time dependence of the quantity $d_H$ appears in Fig.13. Note that a decreasing value of $d_H$ indicates a period of acceleration/inflation since, $d/dt(c/aH) = -\ddot{a}/(\dot{a}^2)$.

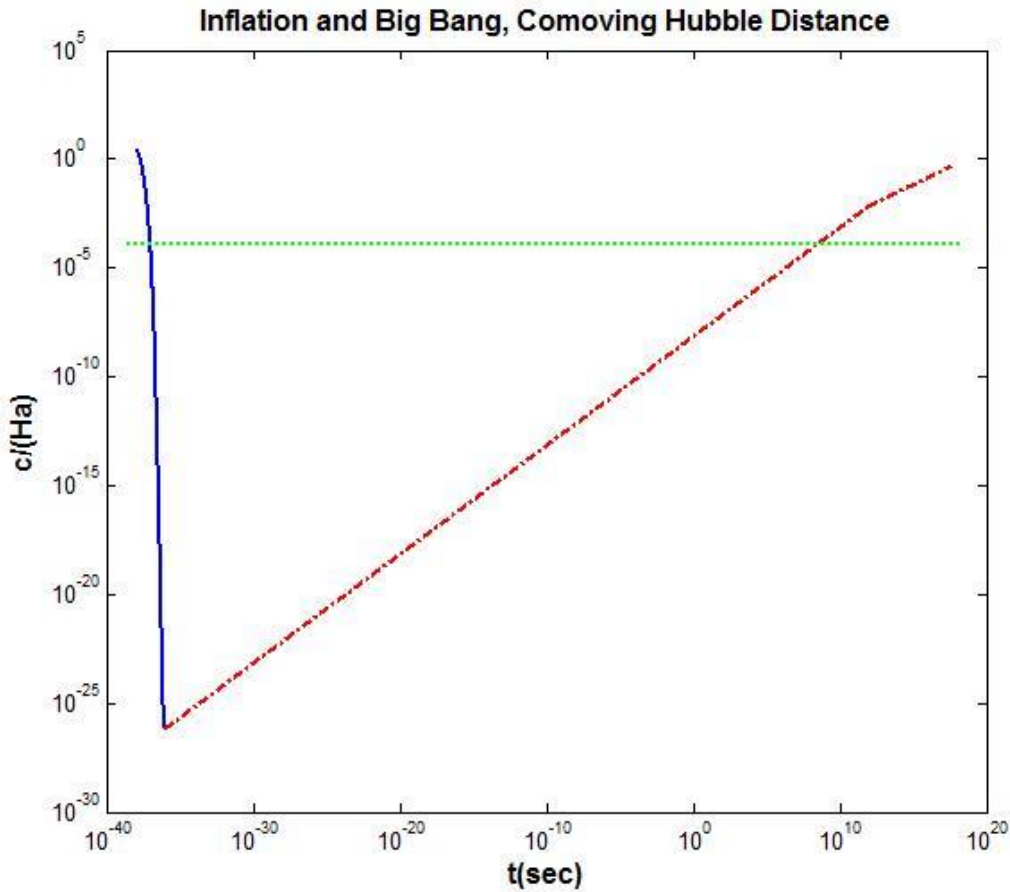

Figure 13: Time dependence of $d_H$ during inflation (blue solid) and a radiation dominated epoch followed by a matter dominated epoch (red dash-dot). The dotted green horizontal line indicates the constant behavior of an arbitrary commoving scale of size 0.001.

As $d_H$ decreases, a horizontal commoving scale will exit the horizon and then re-enter it later during the hot BB phase. For example, a comoving scale of 0.001 would re-enter the horizon before the time of the CMB at $t_{dec}$, while larger comoving scales would re-enter the horizon later. Smaller wavelength perturbations re-enter the horizon earlier and can evolve sooner into gravitational structures than those with larger wavelengths. The observed CMB scales are produced early during inflation. Smaller scales, corresponding to galactic sizes, exit the horizon later in the inflationary period.

Before leaving this section, the somewhat arbitrary nature of the numerology should be mentioned. As stated above, the initial time for inflation and the time span of inflation are chosen, in concert with the scalar mass to give sixty e-folds. These choices are somewhat arbitrary and are meant to be illustrative only. Small changes in this simple model make large effects and the aim is not to achieve a perfect data fit but simply to show that the quartic scalar field model can explain the existing data for BB cosmology.

## 10. Reheating

At the end of inflation, ε = 1 and the energy density ~ V(φ), with $m/M_p = 10^{-6}$, is $\rho_f$ ~ $(6.4 \times 10^{15}$ GeV$)^4$. During reheating the field oscillates about a potential minimum and decays. How this occurs is unknown in any detail, and the degrees of freedom the field couples to are speculative. The reheat temperature could be as low as 1 TeV (supersymmetry or SUSY scale) or as high as $10^{16}$ GeV (grand unified theory scale or GUT scale). Baryo-genesis can be accommodated at the GUT scale, for example.

The temperature scales as 1/a(t) which means the Universe cools by an enormous factor during inflation. This low temperature drives the density of relic states, for example monopoles, to be very low thus inflating away any relic particles to levels consistent with present limits. This situation then evolves into the successful BB epoch, but the dynamics of this evolution is unknown.

The inflationary paradigm is successful and provides testable predictions for the future. However some obvious questions remain. In this note a simple radiation dominated regime has been assumed to be in place immediately after the end of inflation and lasting until $t_{eq}$. This assumption is clearly very simplistic and is made only to be as simple as possible.

At the end of inflation all the energy resides in the field. The decay of the field into SM particles leads to a relativistic fluid. If this fluid is hot enough and forms rapidly enough, the hot BB cosmology is recovered. Inflation needs to end quickly to allow reheating to occur soon enough to preserve the successes of the hot BB cosmology

The scalar field, after the end of the rundown of the potential, acts like a damped oscillator, Eq.14. The remaining potential decays and the Universe is re-heated and thermalized. The equation of motion allowing for the addition of a decay width Γ to Eq.14 is;

$$\ddot{\varphi} + (3H + \Gamma)\dot{\varphi} + m^2\varphi = 0 \qquad (19)$$

In general thermal equilibrium exists if the reactions are not smoothed by the expansion, $\Gamma \gg H$.

Dimensional argument leads to a decay width, Γ, which is proportional to the scalar mass, m, and some coupling constant, α, if the decays of the field are first order two body decays. The decay daughters, x, are assumed to be SM objects or at least objects with masses much less than the scalar mass m [14].

$$\Gamma(\varphi \to xx) \sim \alpha m \qquad (20)$$

Since the Hubble parameter scales as H ~ 1/t in a radiation dominated phase the fraction of the contribution of produced particles to the total energy density is large when 3H decreases at large times and becomes comparable to Γ or $\Gamma^2 \sim (3H)^2 = (72\pi\rho/3M_p^2)$, using Eq.3. Applying the

(21)

Stefan-Boltzmann law, with a number of degrees of freedom, $g_* \sim 100$ appropriate to the SM at high temperature, leads to a thermalized re-heating temperature $T_R$. If all the x particles indicated in Eq.20 are SM then they are all relativistic so that $T^4$ behavior holds;

$$\rho_{re-heat} \sim (\Gamma M_p)^2 / 24\pi = g_* \pi^2 T_R^4 / 30$$

The grand unified models [14], assuming that SUSY intervenes at masses of about one TeV, imply a unified dimensionless coupling constant of about 1/24 active when the strength of the three SM forces come together at about $10^{16}$ GeV or about $10^{-3}$ $M_p$, Fig.14.

Solving Eq.21 for the reheating temperature $T_R = 0.14\sqrt{\Gamma M_p} = 2.8 x 10^{-5} M_p$ which is about a factor 30 larger than the scalar mass m. There is a GUT and m mismatch. Clearly, a better model of reheating is called for in order to reconcile these difficulties and "pre-heating" has been advocated [17].

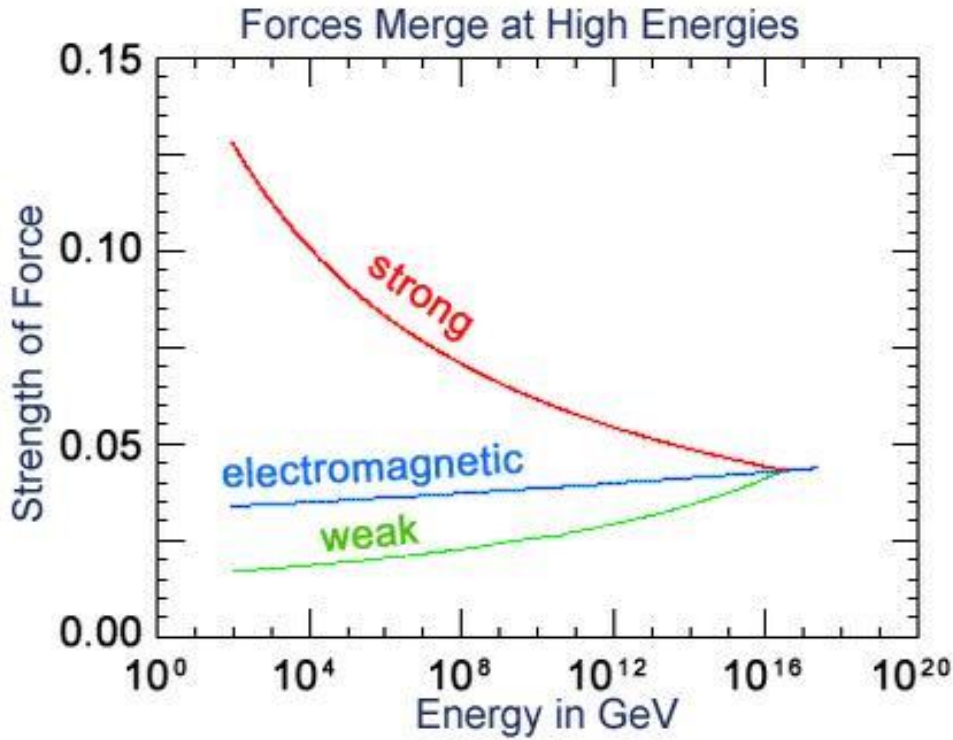

Figure 14: The "running" of the SM coupling constants for the 3 SM forces. They converge at $M_{GUT} \sim 10^{16}$ GeV with a common "fine structure constant" α of about 1/24.

Extrapolating backward in time from $t_{eq} \sim 10^{12}$ sec and $T_{eq} \sim 1$ eV, and scaling for radiation domination to the end of inflation, $10^{-36}$ sec (factor in scale a(t) of $10^{24}$) T rises to $\sim 10^{15}$ GeV or $\sim 10^{-4}$ $M_p$. The simple reheating estimate gives a reheating temperature, $T_R$, 3.4 x $10^{14}$ GeV.

Assuming a radiation dominated epoch, and scaling from $t_{eq}$, the time for the GUT scale is $10^{-34}$ sec.

Clearly a more complex model of reheating is called for [18] which would define the time needed to reheat and the decay modes of the field is the context of a particle physics model relevant at these elevated energies. For the m decays, $H/M_p$ is ~ $10^{-6}$ at the end of inflation and with $m/M_p = 10^{-6}$ the ratio of $3H/\Gamma$ is ~ 71 which means that with $H$ ~ $1/t$, decays will not dominate until a time of $7.1 \times 10^{-35}$ sec. The assumption of an instantaneous reheating is not self-consistent.

Reheating begins with the decays of m into "SM" particles. These particles then decay into the "stable" particles of the SM, neutrinos, photons, electrons, quarks, and dark matter (DM). The quarks are contained, after nucleo-synthesis, in hydrogen, deuterium, lithium and helium. The DM is assumed to be heavy, about 1 TeV in the super-symmetry (SUSY) model. It is therefore "cold" at relevant CMB temperatures and interacts only gravitationally or weakly. It will begin to aggregate gravitationally as soon as it becomes non-relativistic.

The neutrinos are known to have a mass, either by direct end-point measurement of weak SM decays, oscillation experiments or by cosmological limits. The sum of the neutrino masses is less than about 0.3 eV [14]. The neutrinos have a present energy density less than 0.12 GeV/m$^3$, while the DM density is 1.5 GeV/m$^3$ and the baryons have 0.28 GeV/m$^3$. Cosmological evolution can be used to set an upper limit on the sum of the neutrino masses. Neglecting the neutrinos, after nucleo-synthesis the Universe consists of photons, DM, electrons and baryons (light nuclei/atoms). The dark energy (DE) becomes relevant only late in the day (Fig. 2).

## 11. Quantum Fluctuations to Density Perturbations

The inflationary hypothesis solves some outstanding issues in BB cosmology, but at the cost of introducing a scalar field. However, it also makes additional testable predictions. Specifically it predicts the fractional magnitude of density perturbations and a power and spectrum of the temperature perturbations in the CMB which are nearly independent of the scale of those fluctuations. Since H is the only scale during inflation and is ~ constant, a scale invariant spectrum of fluctuations is predicted. Because inflation predicts gravity waves, which are as yet undetected, inflation makes a major prediction.

During inflation there are irreducible zero point quantum fluctuations of the scalar field. These fluctuations are amplified by inflation into classical density perturbations. Although inflation is smoothing out all quantities such as the spatial curvature, quantum fluctuations are intrinsic. The high energies and short times appropriate to inflation imply that the effects of quantum mechanics are not negligible. For fields of strength ~ $M_p$ and times ~ $t_i$, the product is ~ $10^{-19}$ GeV*sec compared to $\hbar$ which is about $10^5$ times smaller which is about the size of the CMB temperature variations. The zero point fluctuations will be stretched by a scale factor ~ $10^{28}$, Fig.9, into classical perturbations. There are also metrical fluctuations due to the gravity of the field which cause both scalar and tensor fluctuations. The subject is complex [17], and only simple order of magnitude arguments are made here.

The commoving Hubble distance, $d_H$, decreases rapidly during inflation, Fig.13. A conformal wavelength $\lambda$ which is initially less than ~ $d_H$ and can be casually active provides the uniformity and small quantum fluctuations ultimately seen in the CMB. The physical wavelength, $a\lambda$, is stretched while the coordinate wavelength $\lambda$ is comoving and constant. During inflation the physical scale goes outside the Hubble horizon $D_H$ and falls out of causal contact, Fig.11. After inflation, $1/d_H$ increases and the wave number crosses the commoving Hubble horizon and becomes causally accessible again. However, due to inflation, these wavelengths are now vastly expanded and are of classical, not quantum size and can serve as the seeds of macroscopic structure formation. The spectrum is defined in units, $k = (Ha) = 1/d_H$. Quantum mechanics plus inflation naturally leads to small but observable density perturbations in the CMB which can be confronted with data.

Quantum fluctuations are formed on sub-horizon scales prior to exiting the horizon. The horizon limits the causal spatial extent of the field, which by the uncertainty principle implies momentum fluctuations. The only scale is H so that the amplitude of perturbations is ~ c/H with energy fluctuation ~ $\hbar H$. The fluctuations, created on a time scale of 1/H, are then estimated to be;

$$\delta\varphi = \hbar(H/2\pi) \tag{22}$$

The factor $\hbar$ is kept explicitly here to highlight the quantum nature of the fluctuation. The field fluctuations cause a local spread in the time of the end of inflation, $\delta t = \delta\varphi/\dot{\varphi}$. A fluctuation in density occurs due to the fluctuation in time, which means the density at the end of inflation varies locally in space. Density fluctuations, $\delta_H$, are therefore created at the end of inflation, $\delta_H \sim \delta\rho/\rho \sim \delta t/t \sim H\delta t \sim H\delta\varphi/\dot{\varphi}$. The dimensionless fractional density fluctuation depends on the square of the scalar field fluctuation and is;

$$\delta_H = \hbar H^2/2\pi\dot{\varphi} = 2\pi(\delta\varphi)^2/\hbar\dot{\varphi} \tag{23}$$

Initially, H is large and the time derivative of the field is small (slow roll). The fluctuation is almost constant and applies to a large range of wavelengths. Therefore, the inflationary model predicts an almost flat spectrum of fluctuations in commoving wave vector.

For a quadratic scalar field, Eq.13, Eq.14, $H^2 = (8\pi/3M_p^2)V(\phi), 3H\dot{\varphi} = -dV/d\varphi$ in the slow roll approximation. A small local value of $dV/d\varphi$ means a longer time delay or a larger density in that region and a larger $\delta_H$. Substituting into Eq. 23 for H;

$$\tag{24}$$

$$\delta_H = \hbar\sqrt{128\pi/3}[V^{3/2}/M_p^3(dV/d\varphi)]$$

The fluctuations depend on both the magnitude and the shape of the potential. In the quadratic scalar field example, $H^2 = 4\pi/3(m\varphi/M_p)^2$, Eq.17 and Eq.18, the fluctuations are;

$$\delta_H = \hbar\sqrt{16\pi/3}[(m/M_p)(\varphi_i/M_p)^2] \tag{25}$$

The dimensionless fractional density fluctuation depends on the square of the initial value of the field and also upon the scalar field mass, m, because m sets the scale for the time rate of decrease of the field, Eq.17.

Gravitational waves also have an amplitude proportional to H during inflation with a power proportional to $H^2$. There is a dimensionless fractional amplitude, $\delta_G$, for such fluctuations [17], $\delta_G = \sqrt{512/3}[V^{1/2}/M_p^2] = 4(H/M_p)/\sqrt{\pi}$, which is analogous to the scalar amplitude $\delta_H$ and can easily be evaluated in the quadratic scalar model. The ratio of $\delta_G$ to $\delta_H$ contains the dimensionless slow roll parameter ε, $1/\varepsilon = 4\pi(\varphi/M_p)^2, \delta_G = \delta_H\sqrt{16\varepsilon}$. Alternatively, for the quadratic scalar model;

$$\delta_G = \sqrt{64/3}[(m/M_p)(\varphi_i/M_p)] \tag{26}$$

Evolution of the perturbations does not occur when they are outside the horizon; they are 'frozen". After inflation the perturbations re-enter the particle horizon and begin to evolve causally. The density perturbations of the photon-baryon plasma lead to CMB temperature perturbations and also seed the evolution of the large scale structure of matter.

Scales which are now observed crossed the comoving horizon near the start of inflation. The fluctuation is evaluated when a physical wave length λa exits the horizon at λ ~ $d_H$ or k = 1/$d_H$. As illustrated in Fig.13, a typical scale, a/1000, re-enters the horizon before the CMB is formed at $t_{dec}$.

The power of the scalar density fluctuations, $P_s$, follows from Eq.24 and Eq.25, is conventionally normalized to be the square of the amplitude and depends on H and ε as;

$$P_s = \delta_H^2 = (H/M_p)^2/\pi\varepsilon = (H^2/2\pi\dot\varphi)^2 \tag{27}$$

A power law factor $[(k/aH)^{n_s-1}]$ with exponent $n_s-1$ is understood to be inserted to allow for deviations from pure scale independence of the power. The scales, (k/$d_H$) have a spectral index $n_s-1$ slightly different from zero. In the case of the quadratic scalar field the perturbation power has a factor with $\varepsilon_i = 1/4\pi(M_p/\varphi_i)^2$, which explicitly labels the parameter ε to be evaluated at the start of inflation;

(28)

$$P_s \sim (16\pi/3)[(m/M_p)^2(\varphi_i/M_p)^4(k/aH)^{n_s-1}]$$

The magnitude of the power goes as the square of (m/$M_p$) and as the fourth power of ($\varphi_i$/$M_p$). The scalar and tensor power can be cast into several forms;

$$P_s = 8(V/\varepsilon)/3M_p^4 = 2/3(m/M_p)^2 N$$
$$P_t = P_s(16\varepsilon) = 48/\pi(H/M_p)^2 \qquad (29)$$

The CMB temperature perturbations occur because photons lose energy climbing out of the gravitational potential of the higher density regions. The CMB fractional temperature anisotropies [14] are about a part in $10^5$ or $\delta_H \sim 10^{-5}$. Specifically, $P_s$ = 2.2 x $10^{-9}$, so that V/$\varepsilon$ = 8.2 x $10^{-10}$ $M_p^4$ = $(0.0054)^4 M_p^4$. In the previous numerical example $\varepsilon$ is 7.5 x $10^{-3}$ and $\varphi_i/M_p$ = 3.25. The required value for m, using Eq.29, is 1.1 x $10^{-6}$ Mp. Indeed, this value for m was chosen to agree with the measured CMB scalar power.

For the spectral index, deviations from scale invariance depend on the shapes of the potential that drives the inflation. A constant value for H would give $n_s$ = 1. The first few terms in the Taylor expansion for the field are contained in the slow roll parameters, $\varepsilon$ and $\eta$. From Eq.28 the index is the logarithmic derivative of the power, $n_s - 1 = d(\ln(\delta_H^2))/d(\ln k)$. When the k scale crosses the horizon, k~1/(Ha)~$e^N$, and using Eq.16 for N, $d(\ln k)/d\varphi = 8\pi/[M_p^2(dV/d\varphi)/V]$. The spectral index is then predicted to be;

$$n_s - 1 = M_p^2/8\pi[-3(dV/d\varphi)^2 + (d^2V/d^2\varphi)] \qquad (30)$$

The scalar index is $n_s$-1 = -6$\varepsilon_i$ + 2$\eta_i$. Using the definitions for the quartic scalar model, $\varepsilon = \eta = (1/4\pi)(M_p/\varphi)^2$ and using the $\varphi_i/M_p$ value of 3.25 the slow roll parameters are both 0.0075 which means a predicted spectral index of 0.969. This is close to the experimental CMB result of 0.948 [14] which has an error of about 0.007. The prediction for the ratio of tensor to scalar power is r = $(\delta_G/\delta_H)^2$ = 16$\varepsilon$ = 0.12. This prediction is in fair agreement with the present upper limit for r.

There are then three basic predictions of the inflationary paradigm; the magnitude of the CMB temperature perturbations, the spectral index of the scalar perturbations, and the ratio of the magnitudes of the tensor to scalar perturbations.

## 12. The CMB Anisotropies

After reheating the Universe evolves in a fashion defined by BB cosmology. The photon – baryon plasma expands and cools until the photons decouple. The CBM decoupling time $t_{dec}$ is about $3.6 \times 10^5$ years. The scale, a(t), has since grown by a factor of about 1100. That this scaled time is about 1/50 of the present Hubble time of about $1.4 \times 10^{10}$ yr, is just another statement of the causal problem in BB cosmology.

The standard CMB analysis expands in spherical harmonics. Two regions of the sky separated by an angle $\Theta$ correspond to a correlation multipole l with $\Theta \sim \pi/l$. The co-moving horizon ratio at $t_{dec}$ and $t_o$ scales at $t^{1/3}$ (Eq.9 and Table2) for matter domination. That ratio is $\sim 0.01$ or about 1 degree of subtended angle at present. More exactly the numerical integral of conformal time shown previously had 0.062 and 2.96 for $\tau$ at $t_{dec}$ and $t_o$. The ratio is 0.021 or $\sim 1.2^o$ so that the horizon for the CMB is twice that or $2.4^0$ or $l = 75$.

The correlation of the CMB temperatures for different angular scales [14] is shown in Fig.15. There is a peak at $l \sim 200$ corresponding to $\sim 1$ degree on the sky which is near to but less than the size of the horizon at decoupling. These and higher l perturbations can have evolved since re-entering the horizon before the CMB formation at $t_{dec}$. These small scales then evolve and begin to experience photon-baryon acoustic oscillations. Low l fluctuations, $l < 75$, were not yet within the horizon at recombination yet the correlations do not vanish, which is only possible because of inflation. These large scales have not evolved from the point where they exited the horizon and are therefore featureless.

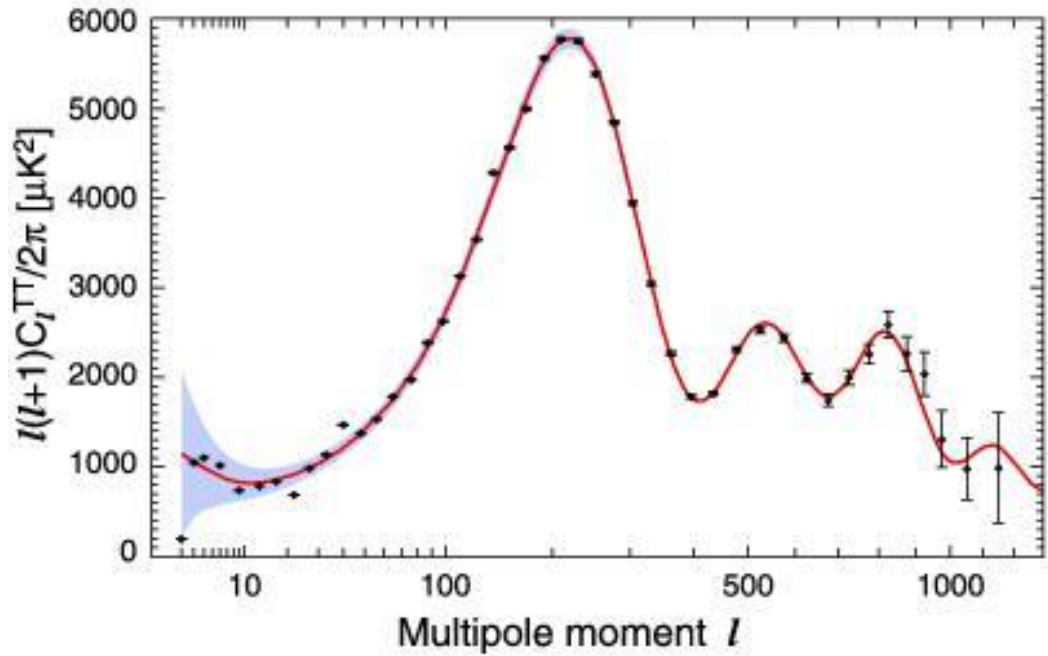

Figure 15: Power spectrum of the CMB temperature perturbations. Structure in the spectrum indicates a phase coherence in the Fourier components, k/(Ha), of the perturbations. The large errors at low l are statistical and irreducible. At large angular scales there are only a finite number of l values that contribute because there are only a finite number of wave modes that fit inside the horizon.

The baryons and photons tend to fall into the potentials of the dark matter which has already begun to form structures since it has not been inhibited by interactions with the photons. Because dark matter has no electromagnetic interactions, it will start to evolve gravitationally as soon as it is pressure-less, perhaps at an energy ~ 1 TeV in some models. Before that, all structures are erased by the dark matter pressure, damping the very small scales which are within the horizon at that time. The distinct regions of the CMB are indicated schematically in Fig.16.

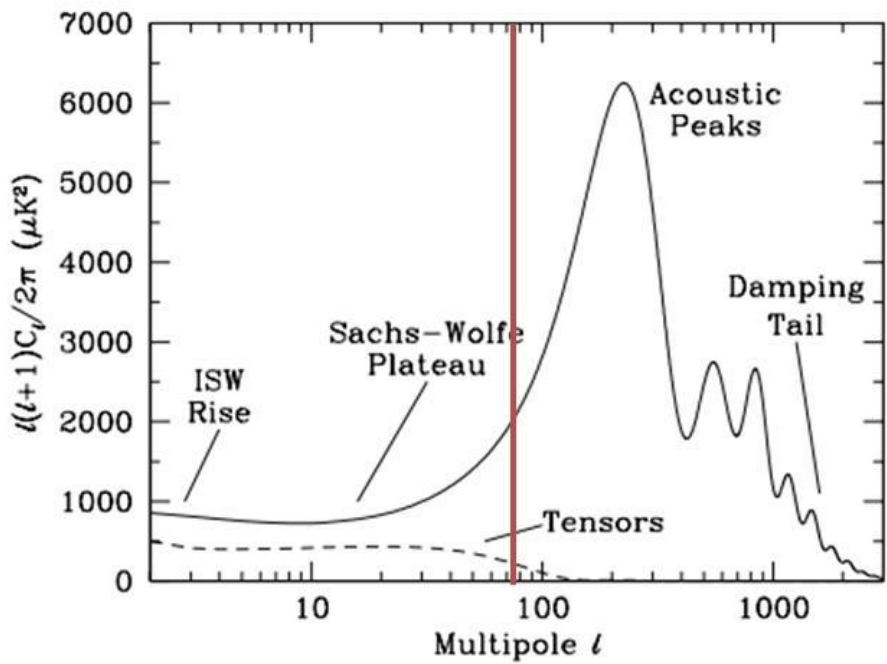

Figure 16: Schematic of the different regions of perturbation size in the CMB. The tensor power is arbitrarily normalized. The horizon is indicated by the vertical red line. Below the horizon the spectrum is featureless, while above it there are several harmonics of the fundamental acoustic oscillation.

It is striking that there are strong oscillatory structures in the CMB spectrum. They arise because the density perturbations evolve under gravity and photon pressure in the strongly coupled photon-baryon fluid. There are acoustic oscillations of the fluid because photon pressure provides a restoring force opposing baryonic gravity which favors gravitational collapse.

Fluctuations at different wavelengths are imprinted on the CMB at different phases in their oscillation. These oscillations can appear in the CMB because the Fourier components of the spectrum all coherently exit the horizon at almost the same time which is a prediction of inflation. Since all Fourier modes with a given wavelength have the same phase they interfere coherently, yielding peaks and valleys in the CMB. Without coherence the spectrum would be featureless. The creation of the CMB at decoupling/recombination is approximately instantaneous which preserves the relative phases of the Fourier components.

The main peak in the temperature correlations of the photons corresponds to a fundamental oscillation mode of the photon - baryon fluid occurring at the speed of sound in the fluid. There are multiple peaks due to the higher harmonics. The next peak is of smaller amplitude since the baryons oppose this mode because it is in a phase of expansion. Higher harmonics are damped because the fluid is not perfectly coupled, baryon to photon, so that there is an added diffusion term causing weaker oscillations. The peak height depends on the baryon and DM abundances and helps to determine their magnitude since more matter damps the oscillations.

After recombination the photons are free and retain the imprint of these oscillations in the temperature anisotropies of the CMB. The baryons will continue to evolve gravitationally into the large scale structures of matter. However, they retain within their structures a memory of the acoustic oscillation. If there is a single field driving inflation, the perturbations for all components; photons, baryons and DM will be in phase. The inflationary model predicts all these features of the CMB remarkably well.

There is another prediction made by inflation for tensor perturbations due to gravity waves which has not yet been tested. The spectral index is $n_G = -2\varepsilon_i$. The power spectrum is expected to be featureless, Fig.16, since there is no interaction with the photons. Quantum fluctuations make only scalar perturbations while gravitational waves make both scalar and tensor perturbations. The power ratio is, $r = P_t/P_s = 16\varepsilon_i$ or 0.12 in the quadratic scalar model.

Current limits on the scalar spectral index and on the tensor to scalar power ratio are shown [12] in Fig.17. The quadratic scalar model prediction is, in fact, rather too close to the present limits and more data will arrive very soon. Tensor fluctuation magnitudes are now only limits since definitive tensor detection is not in hand. It is an exciting time for the inflationary paradigm since the shape of the inflationary potential will soon begin to be restricted. The quadratic scalar results; ns = 0.969 and r = 0.12 with N = 64.6 are close to the large black dot in Fig.17 obtained by much more sophisticated calculations. One dimensional measurements are $n_s$ = 0.948 +- 0.007 and r < 0.11 [14]. Other, models give a range of possible values in the ($n_s$,r) plane. The ratio r is directly proportional to ε.

The CMB is predicted to be polarized due to Thompson scattering of the electrons and photons. The polarization magnitude is related to the temperature perturbations. Specifically, there are so-called "B modes" which are rotational or curl modes of the polarization which cannot be excited by scalar perturbations. Therefore, detection of a B mode [10], [11] polarization of the CMB perturbations would be another confirmed prediction of inflation. The present experimental situation for B modes is not clear [10], [11]. The E mode or divergence modes have been observed and the "E-E" correlation of E mode polarization has been measured. The data is shown in Fig.18 where the acoustic peaks are out of phase with the scalar temperature peaks because the polarization arises from the scattering which is related to the fluid velocity and not the density.

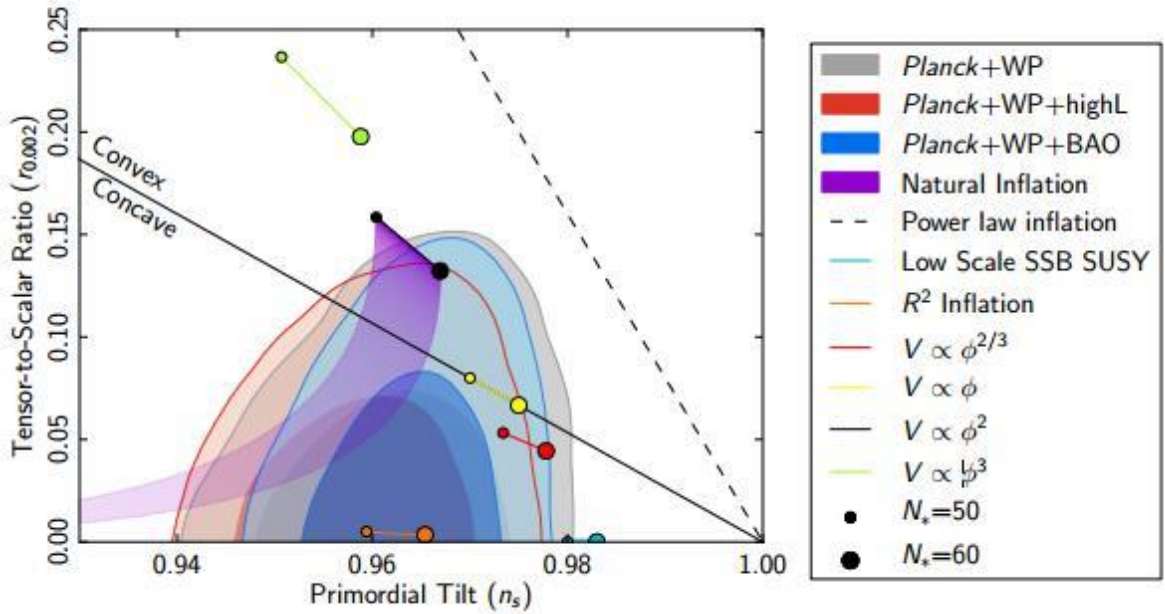

Figure 17: Constraints on $n_s$ and r at present. Also shown are the predictions of several inflationary models. The quadratic scalar model is consistent with, but near to, the present limits.

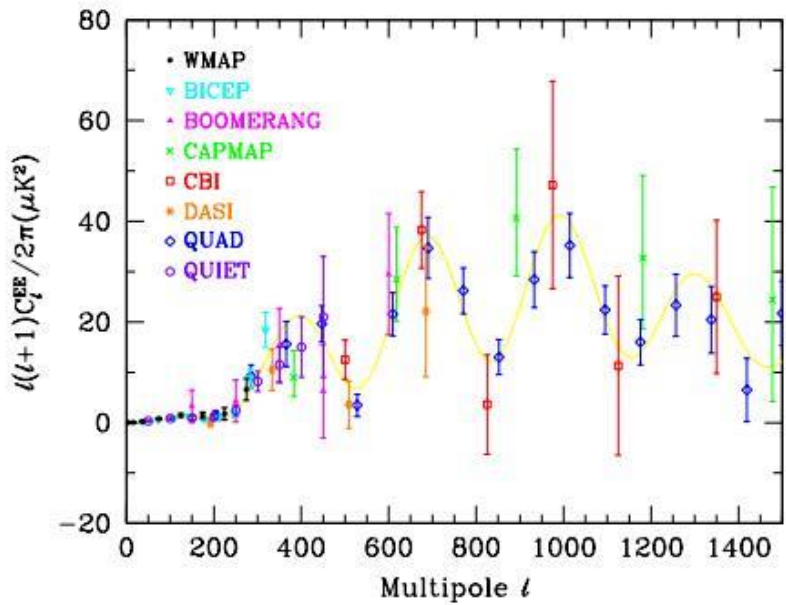

Figure 18: Power spectrum for the E polarization modes analogous to the temperature, T, perturbation. Acoustic oscillations are evident and more data will appear rapidly.

Next generation experiments aim to make definitive measurements of the "B-B" modes which are solely due to gravitational waves. A sample of the predicted results is shown in Fig. 19. If the r value is large a definitive measurement will be made.

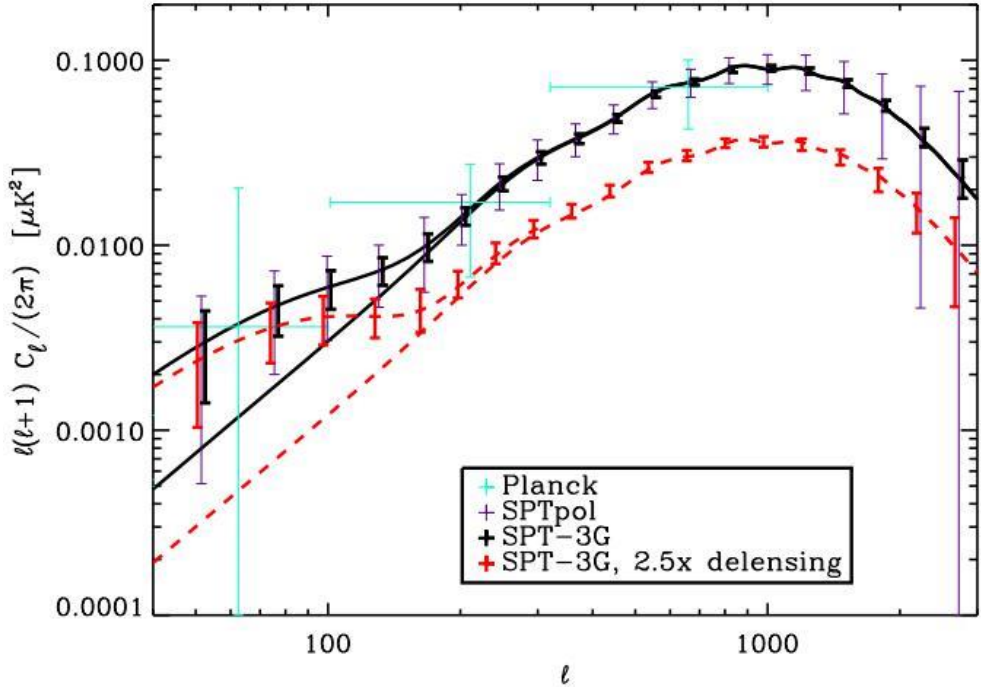

Figure 19: Tensor B mode sensitivity planned for next generation South Polar Telescope (SPT) experiment as a function of multipole l value.

## 13. Evolution of the Structure of Matter

The quantum zero point fluctuations exiting the horizon during inflation, $\sim H^2/\dot{\phi}$, become density perturbations entering the horizon afterwards, dρ/ρ, which are imprinted on the CMB as temperature anisotropies, dT/T and which seed the structure of the Universe today. The comoving scale is k = Ha = $1/d_H$ while the physical scale is K = $1/ad_H$ = $1/D_H$. The preferred unit of length for structure data is the parsec which is about 3.26 light years (lyr) or 3.08 x $10^{16}$ m. The typical distance between galaxies is ~ $10^6$ lyr. $D_H$ is about 4400 times larger, containing about 8x$10^{10}$ galaxies. The distance between galaxies is about 1 Mpc while it is ~ 10 Mpc between clusters of galaxies with the largest clusters of size about 100 Mpc.

The density perturbations at $t_{dec}$ serve as seeds of structure formation because the baryons then evolve independently of the photons and that evolution is gravitationally unstable. Once the DM is inside the horizon it begins to evolve. There is a fairly rapid collapse once the DM becomes non-relativistic because the dark matter does not interact with the photons and hence has no photon pressure opposing the gravitational attraction. The structural evolution of baryonic matter is suppressed until decoupling because of photon pressure. Pressure overcomes gravitational clustering for wavelengths smaller than the horizon during the radiation dominated epoch before $t_{dec}$.

The fractional temperature anisotropy of the photons remains constant after decoupling at the CMB value of about $10^{-5}$. A baryon density perturbation of about the same size grows after $t_{dec}$ until the present. A fractional baryon density perturbation has evolved to about 0.1% today and is roughly compatible with the large scale galactic structures which are observed.

By itself gravity is unstable and a density perturbation will grow without a pressure gradient to stabilize it [19]. Ignoring pressure, for a static Universe, the density perturbation δ will grow exponentially with a time constant $t_{col} = M_p/\sqrt{4\pi\rho}$. This growth is resisted by pressure gradients which move with the speed of sound in the medium, $c_s$. The size of an object over which the pressure may respond to stave off the unstable growth is the Jeans length $j_{ob} = 2\pi c_s t_{col}$. Objects of larger size will collapse if over-dense while smaller objects will oscillate in density as stable sound waves.

When expansion is taken into account, with an expansion time of 1/H, the Jeans scale is comparable to the collapse time for any density which means the two processes are comparable in their influence. In particular, for radiation, with sound speed $c_s = c/\sqrt{3}$, the Jeans length is comparable to the Hubble length;

$$(\lambda_J)_\gamma = 2\pi\sqrt{2}(c/H)/3 \qquad (31)$$

Since this is about 3 times the Hubble distance, density perturbations in radiation will be pressure supported as stable sound waves. Therefore the CMB perturbations will remain stable after decoupling and are observed essentially as first imprinted on the CMB.

For the baryons, before decoupling there is a photon-baryon fluid which oscillates. The Jeans length is about 1.3 x $10^{22}$ m at $t_{dec}$ or about 1/10 of $D_H$. After decoupling the photons and the atoms propagate separately and there is a great drop in pressure for the baryons, since the sound speed for the baryons becomes ~ $c\sqrt{kT/M_{Hy}c^2}$. Since kT is about 0.32 eV at decoupling and the mass for hydrogen, $M_{Hy}$, is about 1 GeV, the Jeans length drops by a factor about 1.5 x $10^{-5}$ after the photons decouple. Before that the tightly coupled matter and radiation both have about the same density perturbations.

After decoupling there is a competition between the expansion which damps out density perturbations and gravity which enhances density perturbations. For the baryon component of the matter, the equation for the perturbation δ has an expansion "frictional term" and a gravitational term proportional to the matter density.

$$\ddot{\delta} + 2H\dot{\delta} - (3\Omega_{m,o}H^2/2)\delta = 0 \qquad (32)$$

The last term is a generalization of the static term for matter alone, $\delta/t_{col}^2$, where $\rho = \Omega\rho_c$ and $\rho_c/H^2$ is a constant. A plot of the densities of matter, radiation and vacuum energy is shown in Fig.20. For the flat Universe the total density is always the critical density.

The results follow from a numerical solution of the differential equation for the scale a(t). For times between the present and $t_{dec}$ a constant value for the matter is a reasonable approximation. In the future, dark energy begins to dominate and the matter density decreases rapidly. The crossover between matter and radiation occurs at $t_{eq}$. For evaluation of H, H = n/t is used in Eq.32. For $\Omega_m$, since $\rho t^2$ is a constant and $\rho(t)$ scales as $1/a(t)^3$ and a(t) scales as $t^{2/3}$, the fraction of matter is approximately constant in a matter dominated epoch. In a radiation dominated epoch the scale a(t) has a time dependence, so that and the matter fraction decreases, $\Omega_m \sim \Omega_{m,o}\sqrt{t/t_{eq}}$, as radiation begins to dominate. These values were used in a numerical solution of Eq.32.

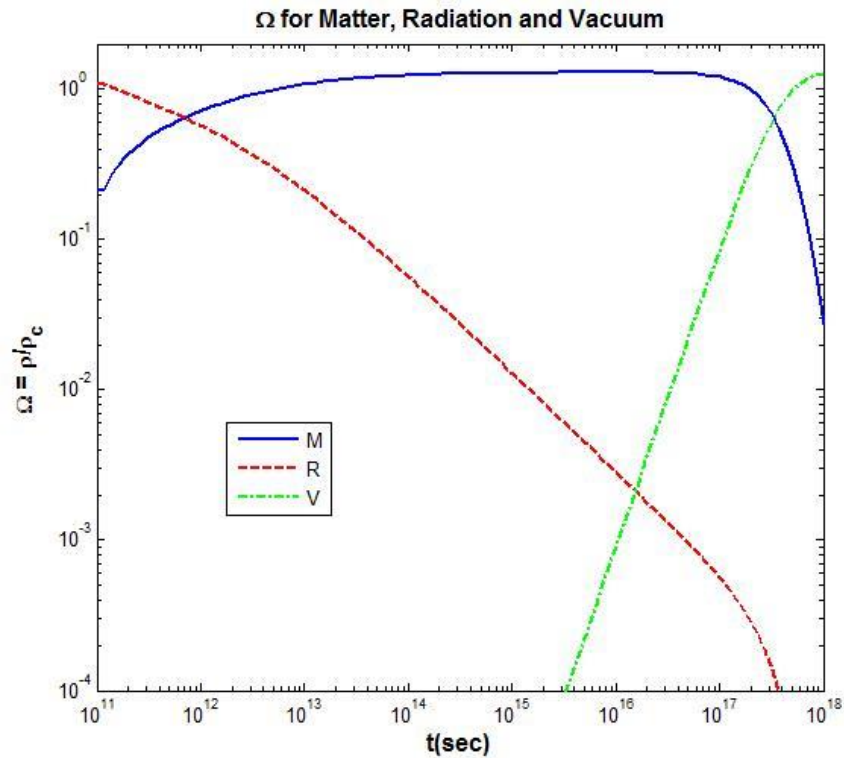

Figure 20: Time dependence of the densities of matter, radiation and vacuum energy scaled to the critical energy, defined by H.

If the Universe is not dominated by matter, then the DM density perturbations do not grow rapidly. If matter dominates, after $t_{eq}$, the DM perturbations grow rapidly with time and there follows a collapse into the gravitationally bound structures observed today. Assuming a power law in t for the solution and $H = (2/3)/t$ appropriate for matter domination the solution to Eq.32 which grows with t has an exponent $[-1/3+\sqrt{(1/3)^2+4\beta}]/2 = 0.32$ where $\beta = (2/3)(\Omega_{m,o}) = 0.21$. The time dependence is seen in the numerical solution of Eq. 32 shown in Fig.21.

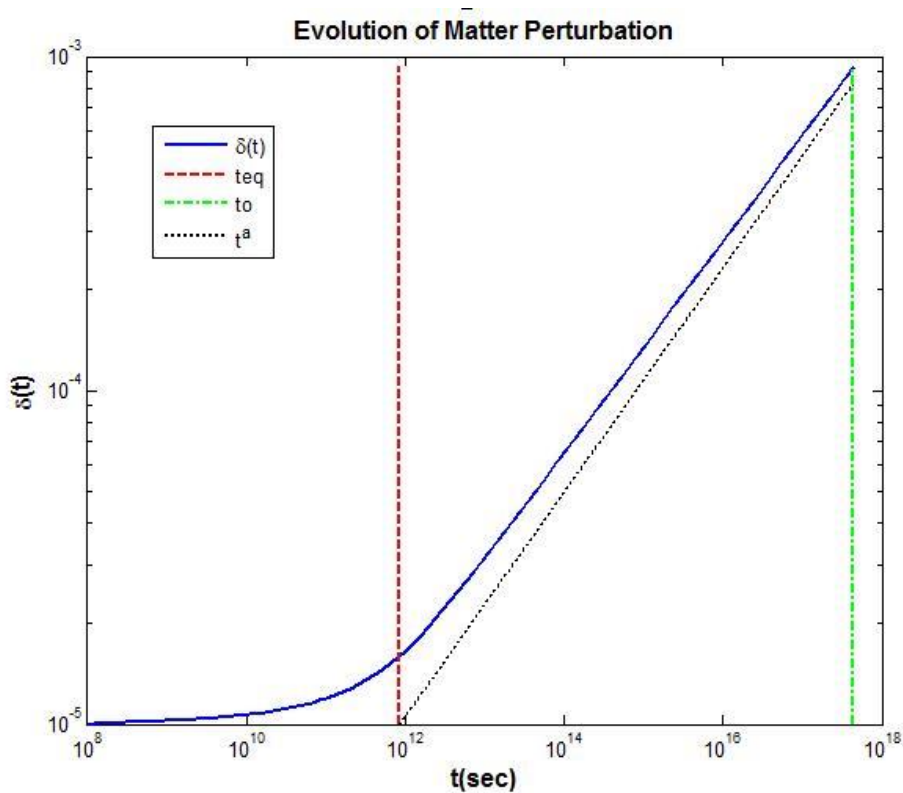

Figure 21: Time dependence of the fractional matter density perturbation (blue-solid). For times before $t_{eq}$ the perturbation is small. After $t_{eq}$ (red, dash) the matter fraction dominates and the perturbation increases rapidly attaining a value of ~ 0.001 at the present time (green-dot/dash). The approximate power law solution appears as a black-dotted line.

The CMB with angular scales from l = 10 to 800 cover K values from ~ 0.0005 to 0.005 $Mpc^{-1}$. Galaxy clusters and their structure can also be used to evaluate the scalar power. They are surveyed at times later than $t_{dec}$ and cover correspondingly greater values of K. The data span a range of K from galaxies to the largest clusters of galaxies. As seen in Eq.28 the basic scale free power has P(k) ~ k at all scales. Data for the CMB, galaxies and clusters appear in Fig.22 assuming the dominance of cold dark matter (CDM).

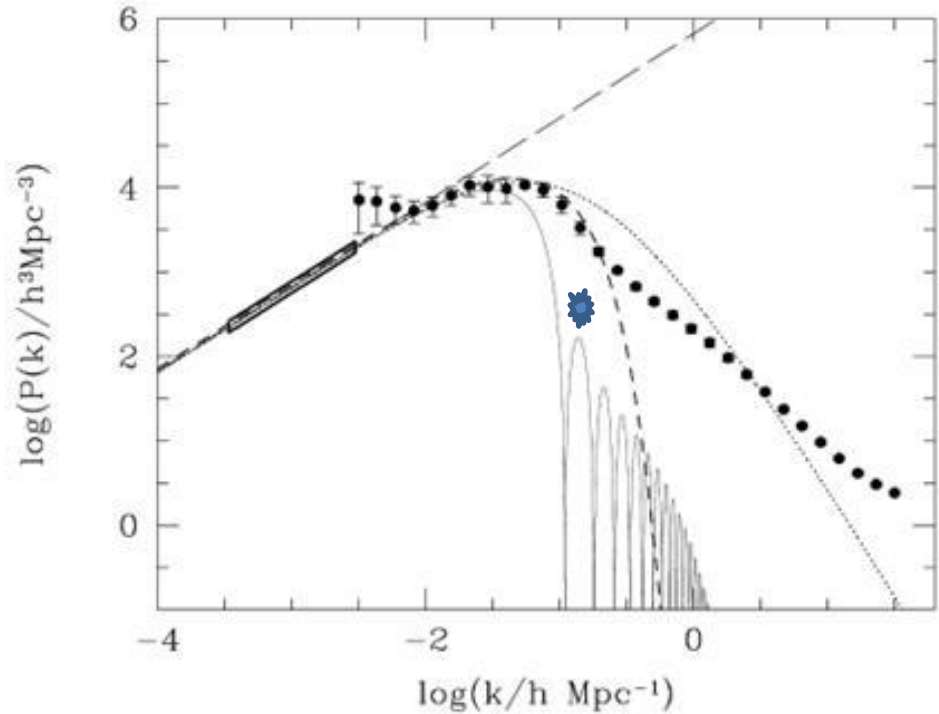

Figure 22: Power spectrum for galaxy clustering. The shaded box at small K represents the CMB data. The long dashed line is for P(k) ~ k, or $n_s = 1$. The dotted, solid and short dashed lines show the spectrum if the critical density is all due to CDM, baryons, or massive neutrinos respectively. The Hubble factor is h ~ 0.73. The star indicates the position of a baryon acoustic oscillations (BAO) peak.

The reduction is power below scale free behavior is best represented by the CDM model provided by massive DM. The effects of baryons and possible massive neutrinos are sub-dominant. The first objects to form are the smallest, so that formation time flows from right to left in Fig.22.

There are small structures in the power spectrum. The evolution of the acoustic waves freezes when the photon pressure ceases. The DM gravitational well causes the baryons to fall in as the structure evolves. The BAO refers to the residual correlations between galaxies as the DM and baryons begin to coalesce. The BAO is the residual oscillation remnant of the acoustic oscillation after the baryons have evolved independently of the photons. Data [20] on the BAO part of the power spectrum appears in Fig.23. The peaks correspond to the structure in the small baryon component seen in Fig.22.

The consistency of the CMB, Fig.15, and the BAO supports the picture of a photon – baryon fluid which freezes at decoupling with the photon structure static, while the baryons evolve under their self- gravitation. The BAO is strong independent confirmation of the inflationary paradigm covering a distinct region of K with respect to the CMB.

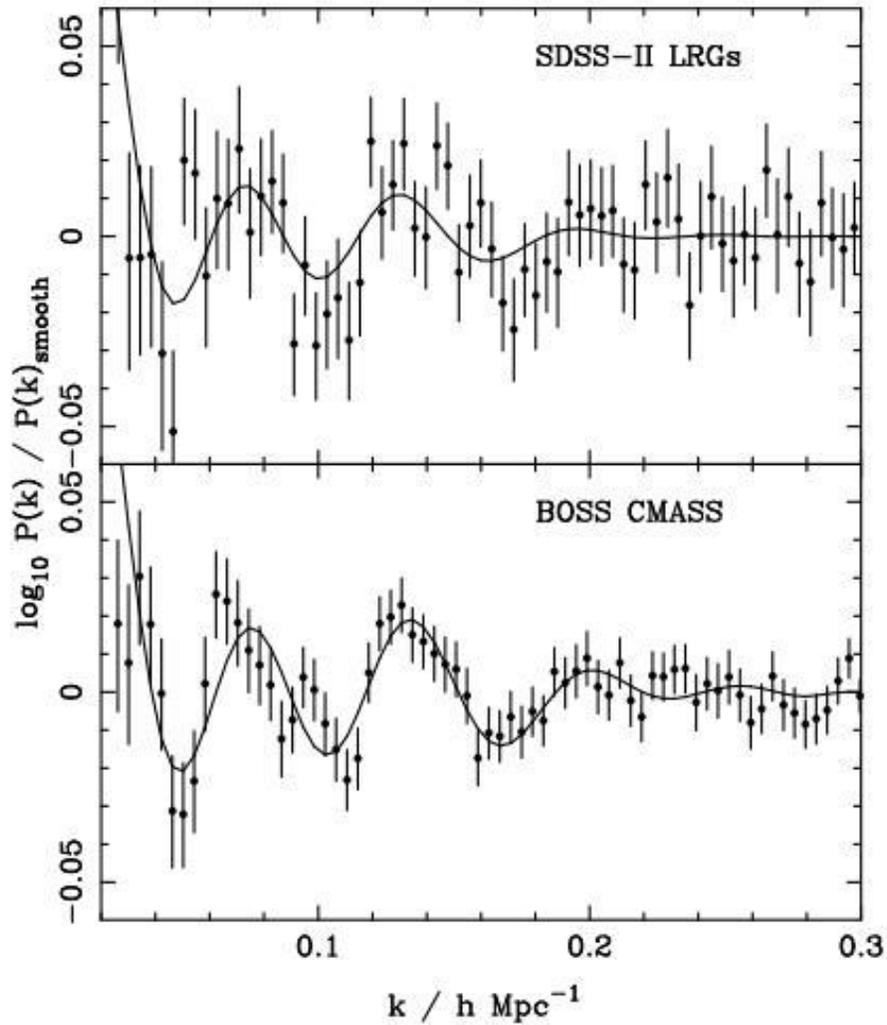

Figure 23: Details of the power spectrum of galaxies for two experiments using redshift surveys of galaxy correlations.

## 14. The Higgs Scalar

A fundamental scalar field has been discovered and inflation requires a scalar field. In the interest of economy there is a possibility that the Higgs boson is the scalar field responsible for inflation. If true, the synergy between particle physics and cosmology would be compelling. This identification is possible if carried out under the assumption that no new physics intervenes between the Higgs scale and the inflation scale. There may, however, be new physics. New physics at the TeV mass scale, for example SUSY, is possible in particular to have a particle physics candidate for DM. New physics at the GUT scale is indicated by the unification of the coupling constants and is often invoked to explain the baryon-antibaryon asyymetry of the Universe. Nevertheless, one can assume this minimal scheme and see where it leads.

The Higgs field is now known [21] to be a scalar, positive parity field, $\varphi$, where P is the momentum, m is the boson mass and S is the action. The Lagrangian density, $\ell$, for the Higgs is shown in Eq. 33. Since the action is dimensionless, the dimension, [ ], of the Higgs field is mass, a specific case of a scalar field.

$$\ell = \bar{\varphi}(P^2 - m^2)\varphi \tag{33}$$
$$S = \int \ell d^4 x, \ [S] = 1, \ [\varphi] = m$$

The self interactions of the Higgs field are specified by defining a potential, Eq.34, which contains two unknown parameters, a quadratic and a quartic self coupling term. The field can have a non zero minimum, giving the Higgs field a expectation value $<\varphi>$ at the potential minimum, which is the vacuum.

$$V = \mu^2 \varphi^2 + \lambda \varphi^4, \ \min \tag{34}$$
$$<\varphi>^2 = -\mu^2 / 2\lambda$$
$$V(<\varphi>) = -\lambda <\varphi>^4$$

The Higgs boson is the excitation of the Higgs field $\phi_H$. Expanding the field about the minimum vacuum value, the potential experenced by the Higgs boson, Eq.35, has four terms.

$$\varphi = <\varphi> + \varphi_H$$
$$V(\varphi) = -\lambda <\varphi>^4 + 4\lambda <\varphi>^2 \varphi_H^2 + 4\lambda <\varphi> \varphi_H^3 + \lambda \varphi_H^4 \tag{35}$$

By inspection, Eq.33, the Higgs acquires a mass, Eq.36 and has triple and quartic self-couplings. There is also a non-zero "cosmological term" which is not relevant at inflationary mass scales. The mass is proportional to the vacuum field value but with an unknown coefficient.

$$m = 2\sqrt{-\lambda} <\varphi> \tag{36}$$

The gauge replacement of the ordinary derivative by the covariant derivative is familiar from classical electrmagnetism. Applying that replacement in Eq.37 to the Higgs Lagrangian kinetic energy term with the electroweak bosons as the covariant fields leads to masses for the W and Z while keeping the photon massless, which was the goal of the Higgs hypothesis. The vector force carriers are represented by $\phi$.

$$\partial \to D = \partial - igV, V = W, Z, \gamma$$
$$D\bar{\varphi}D\varphi \sim (g_W^2/2)\bar{\phi}_W\phi_W(<\varphi>+\varphi_H)(<\varphi>+\varphi_H)$$
$$M_W = g_W <\varphi>/\sqrt{2}, M_Z = M_W/\cos\theta_W, M_\gamma = 0 \quad (37)$$

The vacuum Higgs field gives mass to the W and Z. The Higgs boson then must interact with the W in triple and quartic couplings which are completely specified in the SM. Although it is not strictly necessary, a Yukawa coupling of the Higgs field to the fermions, $\psi$, can be postulated. In analogy to the bosons the fermions then acquire a mass proportional to the Higgs vacuum field and possess coupling to the Higgs boson whose strength is proportional to the fermion's mass. However, this scheme is ad hoc and is not necessary in the SM save for its' appealing simplicity and economy.

Numerically, the Fermi coupling constant, G, is measured in muon decay and the weak coupling constant in neutrino interactions. The data acquired for muon decay, Eq.38, allows for the determination of the Higgs vacuum field. Therefore, there was only one unknown parameter remaining in the SM and that was determined by measuring the Higgs mass.

$$G/\sqrt{2} = g_W^2/8M_W^2, M_W/g_W = <\varphi>/\sqrt{2}$$
$$<\varphi> = \sqrt{2}/4G = 174\,GeV \quad (38)$$

The weak coupling constant, $g_W$, the Weinberg angle relating the electromagnetic coupling and the weak coupling, $\sin\theta_W$, the weak constant, $\alpha_W$, (Fig.14), and the masses predicted for the W and Z, Eq.39, using the vacuum field were spectacularly confirmed at CERN in the 1980's. The Higgs boson mass is not predicted, but 125 GeV was recently [1], [2] measured completely determining the Higgs potential.

$$\sin\theta_W = 0.481, g_W = 0.63, \alpha_W = g_W^2/4\pi = 1/31.6$$
$$M_W = 80\,GeV, M_Z = 91\,GeV$$
$$m = 125\,GeV \to \lambda = 0.60, \mu = 191\,GeV \quad (39)$$

There is one remaining issue, seen in Eq. 35. The vacuum Higgs field is about 7.6 x $10^{13}$ times larger than the observed dark energy. This field appears to be gravitationally inert, which remains an issue to be resolved.

Exploring if it is possible there is no new physics between the Higgs and Planck mass scales it is necessary to evolve or "run" all the couplings as the mass scales changes in accord with quantum field theory. The prescription results in the Renormalization Group Equations (RGE) appropriate for the specific interactions. An example appears in Fig. 14 for the SM couplings. At high mass scales the quadratic term in the potential dominates, so that the running of the λ coupling needs to be explored. The quantum loops contain the Higgs itself due to self-couplings and the strong coupling to the top quark. The top loop pushes the coupling to zero which makes the vacuum, Eq.34, unstable. The Higgs loops make the coupling very large, indicating a breakdown of the theory. For example, the maximum Higgs mass is approximately $4\pi <\phi> / \sqrt{3\ln(M_p^2/2<\phi>^2)}$ for no new physics up to the Planck mass, which sets a maximum Higgs mass of ~ 145 GeV. A more accurate evaluation of the upper and lower Higgs mass limits appears in Fig.24. If the Higgs mass is within a small window near the experimental value a consistent SM theory can be preserved which means that assuming no new physics is at least possible.

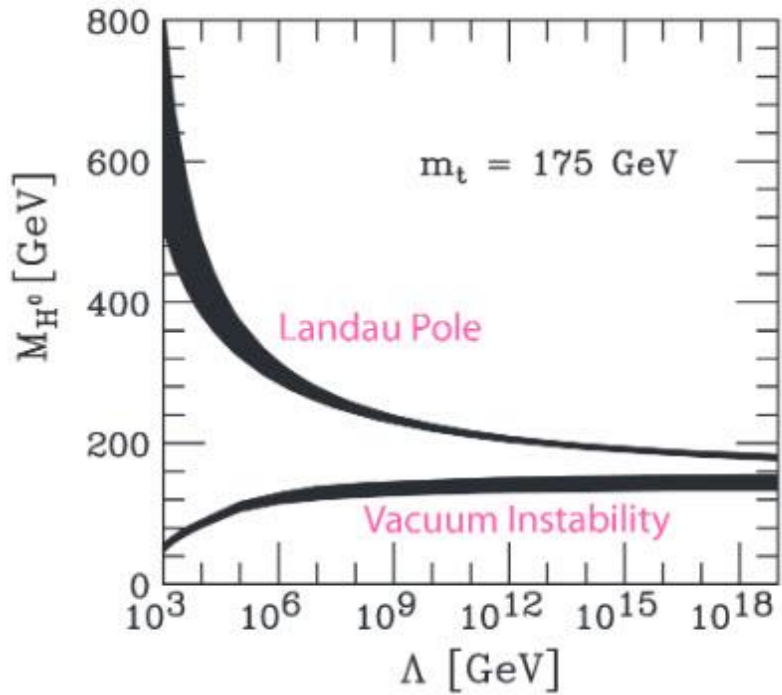

Figure 24: Limits on the Higgs mass due to vacuum instability (top mass) and strong Higgs coupling where λ diverges at mass scale Λ.

## 15. The Higgs as the Scalar of Inflation

The Higgs scalar can be explored at high mass scales as a possible field for inflation. The model should agree with data for the CMB total power and the spectral index of the power. It should be consistent with the limits on r and give a sufficiently large value of N. Assuming measured values for DE and DM, the CMB and BAO oscillations should also be predicted consistently.

A straightforward approach is to use the high mass part of the Higgs potential, Eq.35, when the field is greater than the vacuum expectation value, $\varphi \sim \varphi_H$, $V(\varphi) \sim \lambda \varphi^4$ making the model a quartic potential model. The required strength of any potential to account for the CMB temperature anisotropy $\sim 10^{-5}$ was previously derived, $V(\varphi)/\varepsilon_i = 3/8(P_s M_p^4) \sim (0.0054 M_p)^4$, in Eq.29.

The slow roll parameter is $\varepsilon = M_p^2/16\pi [dV/d\varphi/V]^2 = 1/\pi (M_p/\varphi)^2$ independent of $\lambda$. The spectrum has an index which is assumed to be $n_s = 0.95$ to approximately agree with present measurements, which means $\varphi/M_p = 4.37$. The CMB power requirement defines the value for $\lambda$ since for this potential $V/\varepsilon = \lambda(\varphi^6/M_p^2)/\pi$. There is an immediate problem in that with minimal gravitational coupling of the Higgs to energy, the coupling to match the CMB data requires a value of $\lambda \sim 3.9 \times 10^{-14}$, not the value fixed by the Higgs mass of 0.6. After running the value of $\lambda$ up to a scale $4.4 M_p$ the value might be tuned to the needed small value, but the fine tuning is unattractive at best. Nevertheless, a quartic model for inflation is still possible if the coupling is small, as discussed later.

A solution to this problem for a Higgs model is to specify the unknown coupling of the Higgs to gravity to be non-minimal and to have the Higgs coupled to the Ricci scalar, R, with Lagrangian density, $\ell \sim \varsigma(R\bar{\varphi}\varphi) - V(\varphi)$ and an unknown coupling constant $\varsigma$. This coupling modifies the relationship of the dynamics, the Hubble parameter, to the gravitational sources [22]. The effective potential is;

$$V(\varphi)/(1+8\pi\varsigma\varphi^2/M_p^2)^2 \tag{40}$$

The non-minimal coupling scheme means that the potential at small field values is that appropriate to the SM Higgs, while at high field scales, $\sim M_p$, the effective potential becomes;

$$V(\varphi) \sim \lambda\varphi^4 \to \lambda M_p^4/[(8\pi)^2 \varsigma^2] \tag{41}$$

The mismatch of the Higgs $\lambda$ parameter to the CMB power anisotropy is evaded because the potential at large field values is controlled by the non-minimal coupling parameter. The effective high mass field approaches a constant, $M_p/\sqrt{8\pi\varsigma}$. The modified potential is quite flat at high energies so that there is a region where it has a slow roll behavior driving inflation with a scale invariant spectrum. Using the required effective couplings of $\sim 10^{-13}$ and 0.6 for the Higgs case at high and low mass scales (Eq.38) and equating the two potentials at a field of Mp, the coupling parameter is large, $\xi/\sqrt{\lambda} \sim 47000$. The potentials appear in Fig.25.

With N $\sim$ 60 e-folds needed to solve the BB problems, the Higgs model predicts a scalar power spectrum and a tensor to scalar ratio [21];

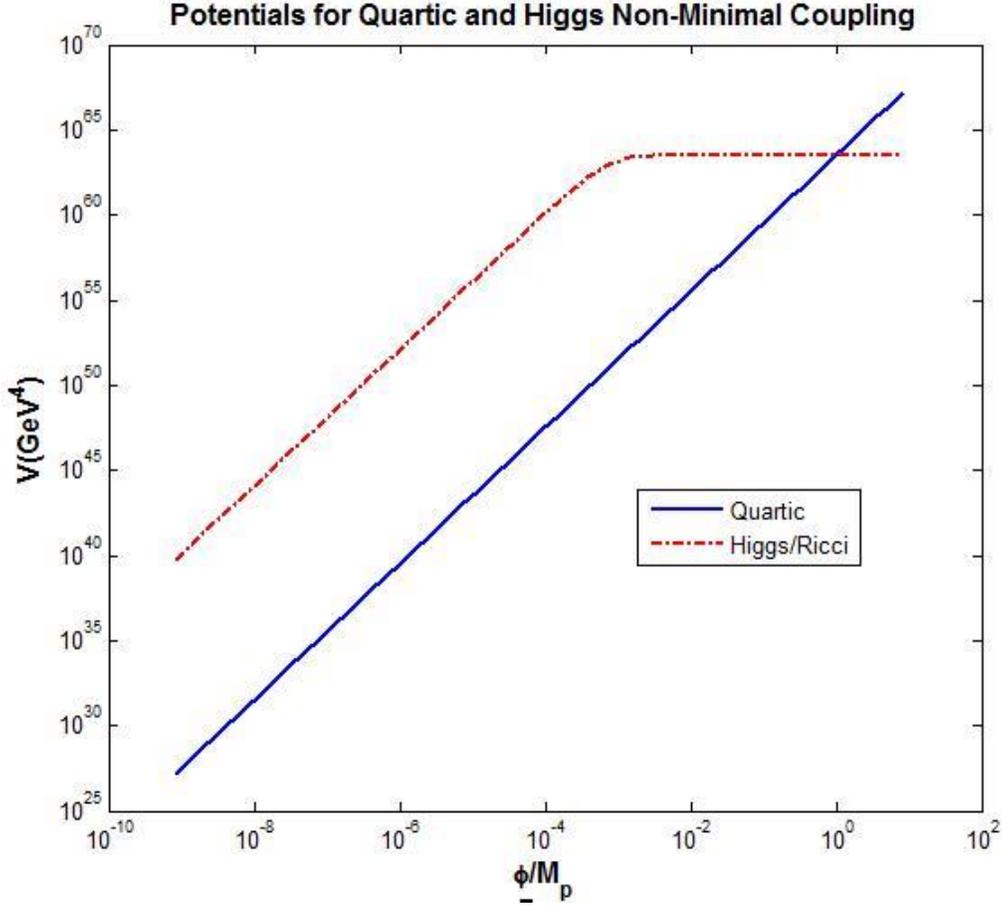

Figure 25: Potentials for inflation for pure quartic coupling to energy and for non-minimal Higgs coupling. The non-minimal case has a very slow roll region where the potential is approximately constant at large values of the field.

$$n_s \sim 1 - 8(4N+9)/(4N+3)^2 \sim 0.97$$
$$r \sim 192/(4N+3)^2 \sim 0.003$$
(42)

For the modified Higgs potential a numerical evaluation of V was used to find N, ε and η. No running of the coupling constants was assumed which makes the results quite approximate. There are 2 parameters, the initial value of the field and the ad hoc coupling constant ς. Because of the flatness of the potential in this case, imposing a value of 60 on N requires a small initial field to $M_p$ ratio of only 0.068. The values of the slow roll parameters are rather low, which drives the r value to be near zero. These parameters are shown in Fig.26 as a function of the field. Inflation ends when ε grows to be one, when φ/$M_p$ is about $10^{-2}$. The 2 parameters predict the values of N, r, $n_s$ and the CMB power. This simple model does not easily predict reasonable values for all 4 observables.

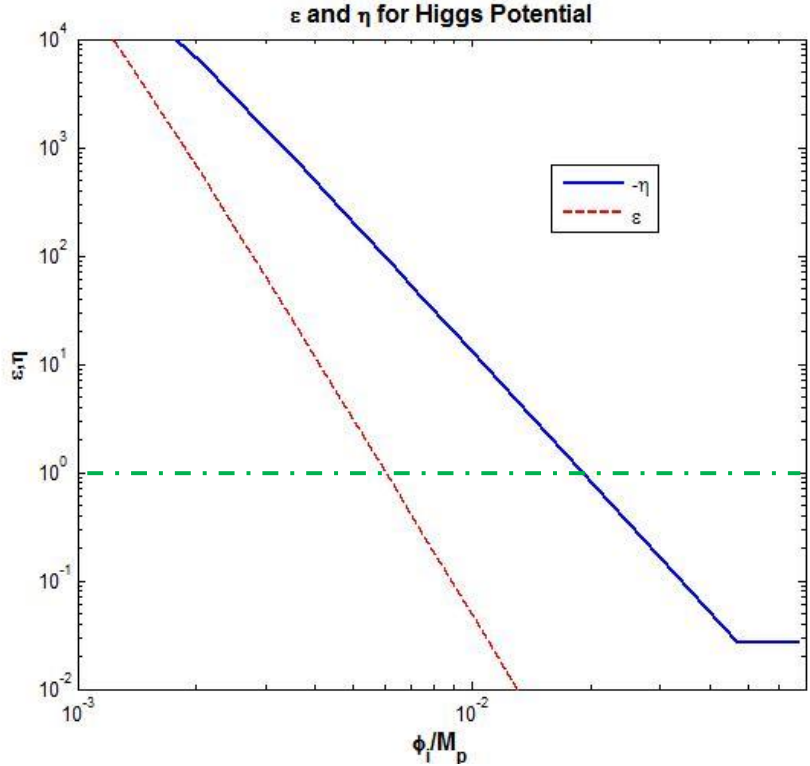

Figure 26: Values of the slow roll parameters for the modified Higgs potential. The value of ε is small, while η is somewhat larger and negative. The dash-dot green line indicates the field value when inflation ends.

The Higgs model has very little tensor power in its' simplest incarnation. Representative values from a more exact calculation are shown in Fig.27. Definitive data on r would begin to distinguish between alternative models for the potential responsible for inflation. The quartic potential model points are large field models and have large r values which are somewhat outside the present experimental limits. The quartic scalar model is presently viable as is the non-minimal Higgs coupling model.

The Higgs couplings, λ(μ) and ς (μ) depend on the energy in the process, μ through the RGE. That dependence, in turn, changes the potential which drives the inflation. A representative plot for the dependencies appears in Fig.28. Without any new physics it is difficult to accommodate precise values of the Higgs and top quark masses. If there is new physics at energy scales above the SM, the RGE would change which would change the predictions of the non-minimal Higgs model. Even so, the simplest and most economical model gives results for non-minimal coupling which are encouraging. Since it is unknown how the Higgs couples to gravity, a coupling to the Ricci scalar cannot be rejected out of hand.

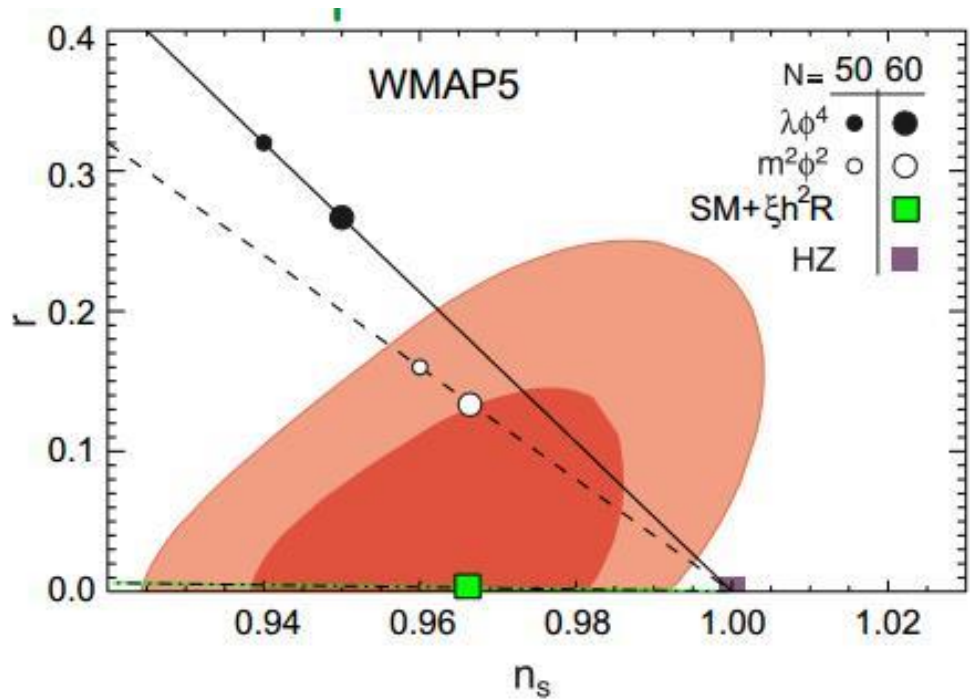

Figure 27: Predicted values of $n_s$ and r for power law and non-minimal Higgs model compared to limits from WMAP. The quartic scalar model is a large field model with large values of r, while the non-minimal coupling Higgs model predicts a small r value.

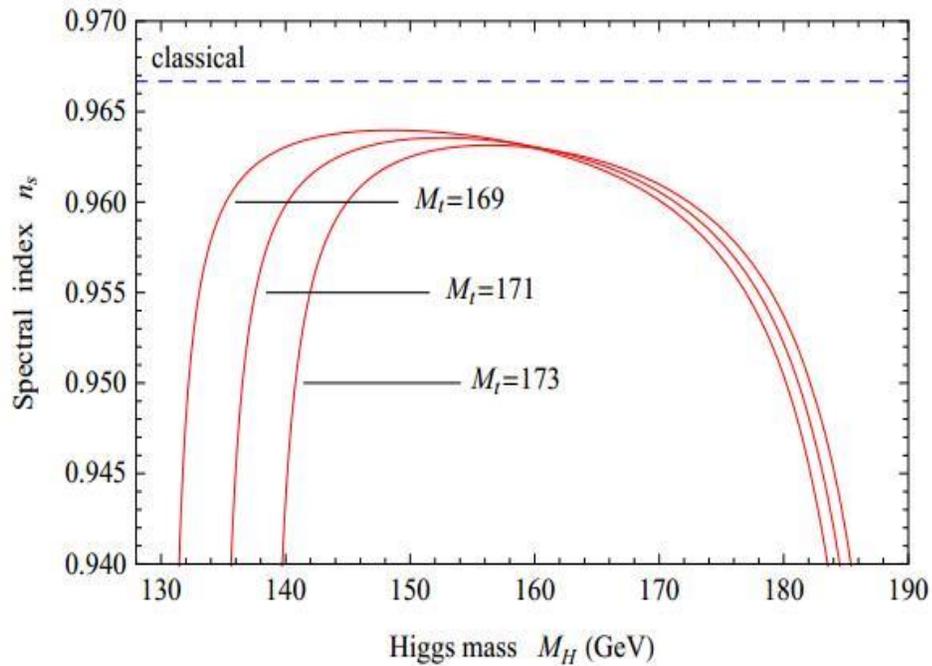

Figure 28: The dependence of the spectral index on the Higgs mass and the top mass. The "classical" line refers to not using the RGE in first or second order and corresponds to the $n_s$ point in Fig.27. There is some difficulty achieving the measured Higgs mass.

# 16. Quadratic, Quartic and Higgs Inflation

Measurements of r and $n_s$ will begin to discriminate between models of inflation in the near future. The differences seen in Fig.27 between quadratic and quartic potentials are a good example, especially combined with the non-minimal Higgs model. A collection of parameters for the two power law models and the modified Higgs model is given in Table3. The values of r and $n_s$ are substantially different once a value of N = 60 e-folds is imposed in order to just solve the issues with BB cosmology.

For the 2 power law potentials there is one free parameter. The numerical coefficients for the slow roll parameters differ in the 2 cases. The value of 1 - $n_s$ is 2/N and 3/N for quartic and quadratic potentials respectively while the r parameter is 8/N and 16/N. The number of e-folds was chosen to be 60 in both cases, which defines the initial value of the field. It also defines the slow roll parameters and therefore the spectral index of the power. The magnitude of the CMB power fixes the coupling constants, m in the quadratic case and $\lambda$ in the quartic case. Finally, the r parameter is defined by $\varepsilon$ alone, while $n_s$ has contributions from both $\varepsilon$ and $\eta$. The numerical results of Table3 are quite consistent with the points in the (r,$n_s$) plane shown in Fig.27. There are small differences from the quadratic and Higgs values quoted in the text because those values arose from a numerical solution of the model.

Table 3: Parameters for Inflation with Quadratic, Quartic and Modified Higgs Potentials

|  | Quadratic | Quartic | Higgs |
|---|---|---|---|
| V | $m^2\varphi^2/2$ | $\lambda\varphi^4$ | Non-minimal |
|  |  |  |  |
| $\varepsilon\pi x^2$, x = ($\varphi/M_p$) | 1/4 | 1 |  |
| $\eta\pi x^2$ | 1/4 | 3/2 |  |
| $N/\pi x^2$ | 2 | 1 |  |
| $V/\pi\varepsilon$ | $2(m/M_p)^2$ | $\lambda x^2$ |  |
| $(1-n_s)\pi x^2$ | 1 | 3 |  |
|  |  |  |  |
| N = 60 -> $\varphi_i/M_p$ | 3.1 | 4.37 | 0.068 |
| 1 - $n_s$ | 0.031 | 0.050 | 0.034 |
| $V/\varepsilon$ = CMB dT/T | $(m/M_p)$ = 1.2 x $10^{-6}$ | $\lambda$ = 3.9 x $10^{-14}$ | $\varsigma$ ~ 77,000 |
| r | 0.13 | 0.267 | ~0.003 |

Improved data will rapidly begin to distinguish between different models for the potential which drives inflation, largely through improved measurements of r and $n_s$.

## 17. Summary

The scalar field responsible for inflation has a value near the Planck mass. The Higgs mass is 125 GeV a factor $10^{17}$ smaller and has a vacuum expectation value of 174 GeV. The other possible scalar field is dark energy with a vacuum expectation value of about 2.3 meV, a factor $\sim 10^{14}$ less than the Higgs field. The connection between these scalar fields, if there is one, is unclear. However, it is now known that one fundamental scalar field exists, so the questions of possible connections and the understanding of the physics operating at early times is fundamental. Indeed, the Higgs scalar can be a possible inflation field if it couples to gravity in a non-minimal manner. These questions are now and for the foreseeable future inaccessible to accelerators, and therefore will likely remain of great importance to both cosmologists and particle physicists.